\documentclass[aps,prd,superscriptaddress,showpacs,nofootinbib,showkeys,10pt]{revtex4-1}
\usepackage{graphicx,subfigure}

\usepackage[usenames]{color}

\def\be{\begin{eqnarray}}
\def\ee{\end{eqnarray}}
\def\suminto{\sum\!\!\!\!\!\!\!\!\!\!\!\!\!\int\limits_{\;\;\;\;\;\{P\}}\!\!\!\!}
\def\sumintoq{\sum\!\!\!\!\!\!\!\!\!\!\!\!\!\int\limits_{\;\;\;\;\;\{Q\}}\!\!\!\!}
\def\sumintob{\sum\!\!\!\!\!\!\!\!\!\int\limits_{P}}
\def\sumint{\sum\!\!\!\!\!\!\!\!\!\!\!\!\!\!\int\limits_{\;\;\;\;\{PQ\}}\!\!\!\!}
\def\sumintb{\sum\!\!\!\!\!\!\!\!\!\!\!\!\!\!\int\limits_{\;\;\;\;P\{Q\}}\!\!\!\! }
\def\nn{\nonumber\\}

\newcommand{\bqa}{\begin{eqnarray}}
\newcommand{\eqa}{\end{eqnarray}}

\begin{document}

\author{Najmul Haque}
\author{Munshi G. Mustafa}
\affiliation{Theory Division, Saha Institute of Nuclear Physics,
Kolkata, India - 700064}
\author{Michael Strickland}
\affiliation{Physics Department, Kent State University, OH 44242 United States}

\title{Two-loop HTL pressure at finite temperature and chemical potential }

\begin{abstract}
\openup 1\jot
We calculate the two-loop pressure of a plasma of quarks and gluons at finite 
temperature and chemical potential using the hard thermal loop perturbation theory (HTLpt) reorganization of 
finite temperature/density quantum
chromodynamics.  The computation utilizes a high temperature expansion through fourth order in the ratio 
of the chemical potential to temperature.  This allows us to reliably access the region of high 
temperature and small chemical potential.  We compare our final result for the leading- and next-to-leading-order 
HTLpt  pressure at finite
temperature and chemical potential with perturbative quantum chromodynamics (QCD) 
calculations and available lattice QCD results.
\end{abstract}

\maketitle

\section{Introduction}

Quantum chromodynamics (QCD) exhibits a rich phase structure and the equation
of state (EOS) which describes the matter
can be characterized by different degrees of freedom depending upon the temperature and
the chemical potential. 
Hadrons are the relevant degrees of freedom at low temperature and chemical potential where chiral 
symmetry is spontaneously broken and quarks and gluons are confined but the matter is approximately 
${\rm SU(3)_c}$ center-symmetric. At high temperatures the system is expected to make a phase 
transition to a quasifree state known as quark-gluon plasma (QGP). In the QGP chiral symmetry is 
restored and the expectation value of the Polyakov Loop becomes close to one, signaling 
deconfinement.\footnote{With dynamical quarks the center symmetry Z(3) in SU(3) is explicitly broken 
yet it can be regarded as an approximate symmetry and the expectation value of the Polyakov Loop is 
still useful as an order parameter.}
At high temperatures
and moderate chemical potentials one therefore expects the system to be in the QGP phase.  Such
conditions are generated in relativistic heavy ion collisions at Brookhaven National Laboratory's
Relativistic Heavy Ion Collider (RHIC)~\cite{rhic}, the European Organization for Nuclear Research's
Large Hadron Collider (LHC)~\cite{lhc}, and are expected to be generated at the Gesellschaft fur 
Schwerionenforschung's Facility for Antiproton and Ion Research (FAIR)~\cite{fair}.

The determination of the equation of state (EOS) of QCD matter is extremely important to QGP 
phenomenology.  There are various effective models (see e.g. \cite{qp,hm,polyakov,njl,pnjl}) to describe 
the EOS of strongly interacting matter; however, one would prefer to utilize systematic first-principles 
QCD methods.  The currently most reliable method for determining the EOS is lattice QCD~\cite{lqcd}.  
At this point in time lattice calculations can be performed at arbitrary temperature, however, they are restricted 
to relatively small chemical potentials~\cite{gavai,fodor}.   Alternatively, perturbative QCD (pQCD)
\cite{lebellac,kapusta,pqcd,kajantie} can be applied at high temperature and/or chemical potentials where 
the strong coupling ($g^2=4\pi \alpha_s$) is small in magnitude and non-perturbative effects are expected 
to be small.  However, due to infrared singularities in the gauge sector, the perturbative expansion 
of the finite-temperature and density QCD partition function breaks down at order $g^6$ requiring 
non-perturbative input albeit through a single numerically computable number~\cite{kajantie,linde}.  
Up to order $g^6\ln(1/g)$ it possible to calculate the necessary coefficients using analytic (resummed) 
perturbation theory.

\begin{figure}[t]
\includegraphics[width=8cm,height=8cm]{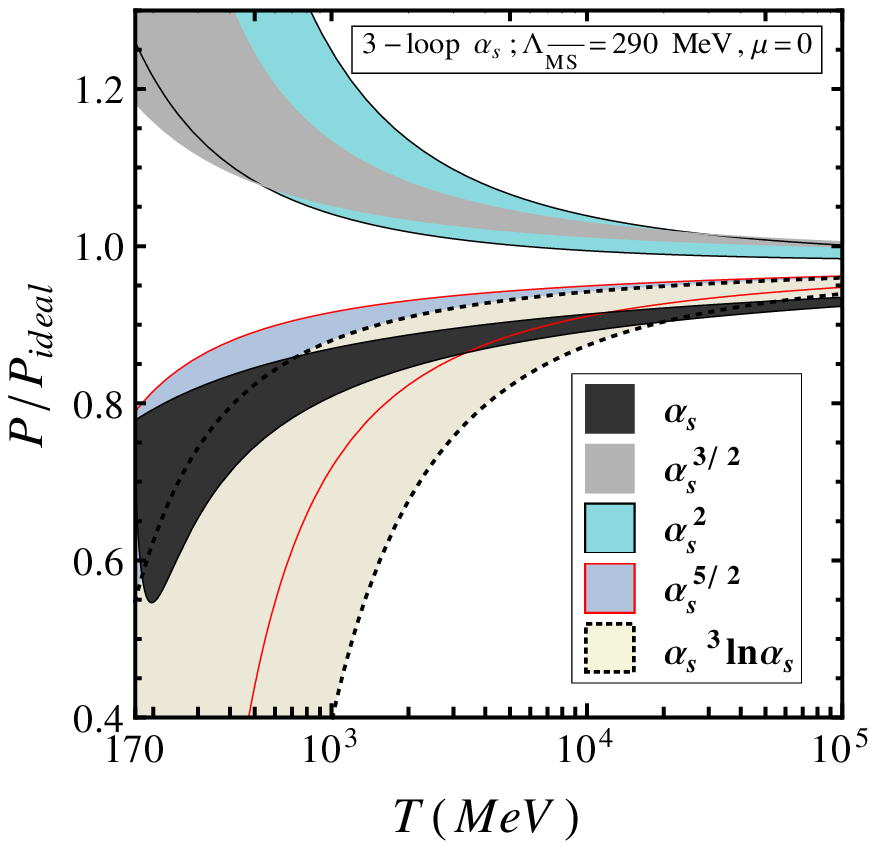}
\includegraphics[width=8cm,height=8cm]{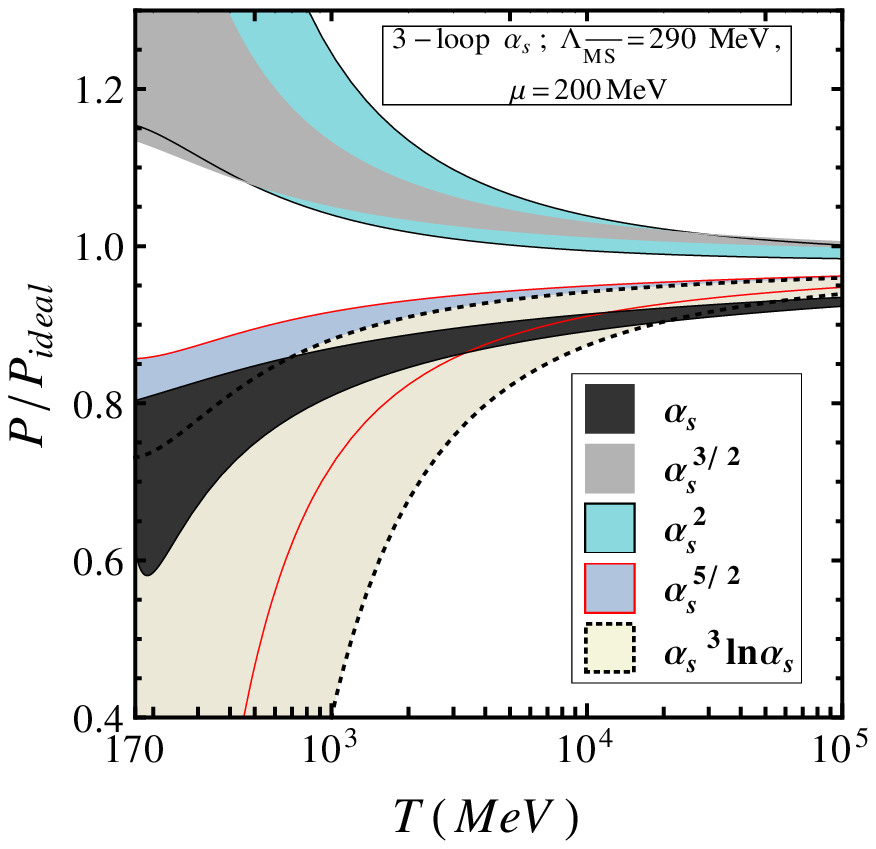}
\caption{The $N_f=3$ pQCD pressure specified in Eq.~(\ref{freeg}) as a function of the temperature.
Successive perturbative approximations are shown through order $\alpha_s^3\ln\alpha_s$ for vanishing $\mu$ (left) 
and for non-vanishing $\mu$ (right).  The shaded bands indicate the variation of the pressure as the $\overline{\rm MS}$
renormalisation scale is varied around a central value of $\Lambda=2\pi \sqrt{T^2+\mu^2/\pi^2}$ \cite{vuorinen,scale} 
by a factor of two.
We use $\Lambda_{\overline{\rm MS}}=290$ MeV based on recent lattice calculations~\cite{latt_lms} 
of the three-loop running of $\alpha_s$.
}
\label{pertfig}
\end{figure}

Since the advent of pQCD there has been a tremendous effort to compute the pressure order by order in the 
weak coupling expansion \cite{pqcd,kajantie,vuorinen,ipp}.  The pressure has been calculated to order of $g^6\ln(1/g)$ at 
zero chemical potential ($\mu=0$) and finite temperature $T$ \cite{kajantie} and finite chemical potential/temperature ($\mu \geq 0$ and $T \ge 0$) \cite{vuorinen}. 
In addition, the pressure is known to order $g^4$ for large $\mu$ and arbitrary $T$ \cite{ipp}.  Unfortunately,
one finds that as successive perturbative orders are included, the series converges poorly and the 
dependence on the renormalization scale increases rather than decreases.     
The resulting perturbative series only becomes convergent at very high temperature ($T\sim 10^5 \, T_c$).
One could be tempted to say that this is due to the largeness of the QCD coupling constant at realistic 
temperatures; however, in practice one finds that the relevant small quantity is, in fact, $\alpha_s/\pi$ 
which for phenomenologically relevant temperatures is on the order of one-tenth.  Instead, one finds that
the coefficients of $\alpha_s/\pi$ are large.  This can be seen by examining the weak coupling expansion of 
the free energy ${\cal F} (T,\mu)$ of QGP calculated~\cite{vuorinen} up to order $\alpha_s^{3}\ln (\alpha_s)$ 
\begin{eqnarray}
{\cal F} &=& - \frac{8 \pi^2}{45} T^4 \,
\biggl[ {\cal F}_0
+ {\cal F}_2  {\alpha_s \over \pi}
+ {\cal F}_3  \left( {\alpha_s\over \pi} \right)^{3/2}
\!\!  + {\cal F}_4  \left( {\alpha_s \over \pi} \right)^2
+ {\cal F}_5  \left( {\alpha_s \over \pi} \right)^{5/2}
\!\! + {\cal F}_6\left( {\alpha_s \over \pi} \right)^3 +\cdots \biggr] , 
\label{freeg}
\end{eqnarray}
where we have specialized to the case $N_c=3$ and
\begin{eqnarray}
{\cal F}_0 &=& 1 + {21\over 32}N_f \left(1+\frac{120}{7}\hat\mu^2 
+\frac{240}{7}\hat\mu^4 \right)  \, ,
\\
{\cal F}_2 &=& - {15 \over 4} \left[ 1 + {5 N_f\over 12}\left(1+\frac{72}{5}\hat\mu^2 
+\frac{144}{5}\hat\mu^4 \right)  \right]\;,
\\
{\cal F}_3 &=& 30 \left[ 1 + \textstyle{1 \over 6}\left(1 + 12\hat\mu^2\right)
N_f \right]^{3/2} \\
 {\cal F}_4 &=&  237.223 + \left(15.963 + 124.773\ \hat\mu^2 -319.849\hat\mu^4 \right) N_f
 \nonumber\\&& \hspace{2mm}
  -\left( 0.415 + 15.926\ \hat\mu^2 + 106.719\ \hat\mu^4\right)N_f^2 \nonumber
 \\ && \hspace{4mm} \nonumber
 + { 135 \over 2} \left[ 1 + \textstyle{1\over 6}\left(1+12\hat\mu^2\right)N_f  \right]
         \log \left[ {\alpha_s \over \pi}
         \left(1 + \textstyle{1\over 6}\left(1+12\hat\mu^2\right)N_f \right) \right] 
 \\ && \hspace{6mm}
 -{165\over 8} \left[1+{5\over12}\left(1+\frac{72}{5}\hat\mu^2 
 +\frac{144}{5}\hat\mu^4 \right) N_f\right]\left(1 -{2\over33} N_f\right)\log{\hat \Lambda}
\; ,
 \\ \nonumber
 {\cal F}_5 &=& -\left( 1 + \frac{1+12\hat\mu^2}{6} N_f\right)^{1/2}
 	\Bigg[ 799.149 + \left(21.963 - 136.33\ \hat \mu^2 + 482.171\ \hat\mu^4 \right)N_f 
 \nonumber\\&& \hspace{2mm}
        + \left(1.926 + 2.0749\ \hat\mu^2 - 172.07\ \hat\mu^4\right) N_f^2\Bigg] 
 \nonumber\\ && \hspace{4mm}
        +\ {495\over 2} \left(1+{1+12\hat \mu^2\over 6} N_f\right)\left(1 -{2\over33} N_f\right)\log{\hat \Lambda}
 \; , \\
  {\cal F}_6 &=& -\Bigg[659.175 + \left(65.888 -341.489\ \hat\mu^2 + 1446.514\ \hat\mu^4\right)N_f 
  \nonumber \\
&& \hspace{2mm} + \left(7.653 + 16.225\ \hat \mu^2 - 516.210\ \hat \mu^4\right) N_f^2
\nonumber\\
&& \hspace{4mm}- \frac{1485}{2}\left(1+\frac{1+12\hat\mu^2}{6}N_f\right)\left(1-\frac{2}{33}N_f\right)\log{\hat\Lambda} \Bigg]
\log\left[\frac{\alpha_s}{\pi}\left(1+\frac{1+12 \hat \mu^2}{6}N_f\right)4\pi^2\right]\nonumber \\
&& \hspace{6mm} -  475.587\log\left[\frac{\alpha_s} {\pi}\ 4\pi^2C_A\right], 
\end{eqnarray}
where here and throughout all hatted quantities are scaled by $2\pi T$, e.g. $\hat\mu = \mu/(2\pi T)$, 
$\Lambda$ is the modified minimum subtraction ($\overline{\rm MS}$)  
renormalisation scale,  and $\alpha_s=\alpha_s(\hat \Lambda)$ is the running coupling. 
At finite $T$ the central value of the renormalisation scale is usually chosen to be $2\pi T$. However,
at finite $T$ and $\mu$  we  use the  central scale $\Lambda = 2 \pi \sqrt{ T^2 + (\mu/\pi)^2}$,  which is 
the geometric mean between $2 \pi T$ and $2 \mu$ \cite{vuorinen,scale}.
In Fig.~\ref{pertfig} we plot the ratio of the pressure to an ideal gas of quarks and gluons.
The figure clearly demonstrates the poor convergence of the naive perturbative series 
and the increasing sensitivity of the result to the renormalisation scale 
as successive orders in the weak coupling expansion are included.  

In this context one should note 
that one can explicitly separate the contributions coming from the soft sector (momenta 
on the order of $g_s T$ where $g_s^2 = 4 \pi \alpha_s$) and the hard sector (momenta 
on the order of $T$) using effective field theory/dimensional reduction methods 
\cite{Braaten:1995cm,Braaten:1995jr,Kajantie:2002wa,Blaizot:2003iq,Andersen:2004fp,Vuorinen:2004rd,Laine:2006cp}.  
After doing this one finds that the hard-sector contributions, which form a power series in even powers of $g_s$, 
converge reasonably well; however, the soft sector perturbative series, which contains odd powers
of $g_s$, is poorly convergent.  This suggests that in order to improve
the convergence of the resulting perturbative approximants one should treat the soft sector
non-perturbatively, or at least resum soft corrections to the pressure.  There have been works
in the framework of dimensional reduction which effectively perform such soft-sector resummations
by not truncating the soft-scale contributions in a power series in $g_s$, see e.g.
\cite{Kajantie:2002wa,Blaizot:2003iq,Laine:2006cp}.  This method seems to improve the
convergence of the perturbation series and provides motivation to find additional analytic methods
to accomplish soft-sector resummations.

In order to better describe the soft-scale contributions there have been various resummation 
schemes developed which attempt to improve the convergence of the successive approximations by 
reorganizing the calculation in terms of quasiparticle degrees of freedom
\cite{resum,braaten,pisarski,andersenx,andersen,%
blaizot,blaizot1,peshier,andersen1,andersen2,andersen3,su,andersen_mu,%
munshi,munshi1,munshi2,jiang}.  These resummation methods include some relevant physical ingredients, e.g.
screening masses and Landau damping.  These reorganizations of perturbation theory canonically include quasiparticle
degrees of freedom from the outset, as opposed to naive perturbation theory.  In the naive perturbative treatment
an expansion around the vacuum is made and one only includes quasiparticle effects in order to regulate infrared divergences. 
Based on Hard Thermal Loop (HTL) resummation~\cite{braaten,pisarski}, a manifestly gauge-invariant reorganization 
of finite temperature/density QCD called HTL perturbation theory (HTLpt) has been developed~\cite{andersen1}.  
HTLpt has so far been applied primarily to the case of finite temperature 
and zero chemical potential. In HTLpt~\cite{andersen1} the next-to-leading order (NLO)~\cite{andersen2} and next-to-next-to-leading order
(NNLO)~\cite{andersen3} thermodynamic functions have been calculated at finite $T$ but $\mu=0$.  Recently ~\cite{munshi1,su} 
the leading order (LO) HTL pressure for finite $T$ and $\mu$ has been calculated and approximately a decade ago it was applied at LO
for finite $\mu$ but $T=0$~\cite{andersen_mu}. 

In view of the ongoing RHIC beam energy scan and planned FAIR experiments,
one is motivated to reliably determine the thermodynamic functions at finite chemical potential.   In this article we compute the NLO pressure of 
quarks and gluons at finite $T$ and $\mu$.  The computation utilizes a high temperature expansion through fourth order in the ratio 
of the chemical potential to temperature.  This allows us to reliably access the region of high 
temperature and small chemical potential.  We compare our final result for the NLO HTLpt  pressure at finite
temperature and chemical potential with state-of-the-art perturbative quantum chromodynamics (QCD) 
calculations and available lattice QCD results.

The paper is organized as follows. In Sec.~\ref{htlpt}, we will briefly review HTLpt. In Sec.~\ref{ingredients} 
we discuss various quantities required to be calculated at finite chemical potential based on prior calculations of
the NLO thermodynamic at zero chemical potential~\cite{andersen2}. In Sec.~\ref{scalarint} we 
reduce the sum of various diagrams to scalar sum-integrals. A high temperature expansion is made in Sec.~\ref{expand} 
to obtain analytic expressions for both the LO and NLO thermodynamic potential.  We then use this to compute the
pressure in Sec.\ref{nlopress}. We conclude in Sec.~\ref{concl}.  Finally, in Appendices~\ref{asis} and~\ref{bsis} we 
collect the various integrals and sum-integrals necessary to obtain the results presented in the main body of the text.  

\section{Hard Thermal Loop Perturbation Theory}
\label{htlpt}

HTL perturbation theory~\cite{andersen1,andersen2,andersen3} is a reorganization of the perturbation
series for hot and dense QCD which has  the following Lagrangian density
\begin{eqnarray}
{\cal L}= \left({\cal L}_{\rm QCD}
+ {\cal L}_{\rm HTL} \right) \Big|_{g \to \sqrt{\delta} g}
+ \Delta{\cal L}_{\rm HTL} \, ,
\label{L-HTLQCD}
\end{eqnarray}
where $\Delta{\cal L}_{\rm HTL}$ collects all necessary renormalization counterterms and 
${\cal L}_{\rm HTL}$ is the HTL effective Lagrangian~\cite{braaten,pisarski}.  It can be
written compactly as
\begin{eqnarray}
\label{L-HTL}
{\cal L}_{\rm HTL}=-{1\over2}(1-\delta)m_D^2 {\rm Tr}
\left(G_{\mu\alpha}\left\langle {y^{\alpha}y^{\beta}\over(y\cdot D)^2}
	\right\rangle_{\!\!y}G^{\mu}_{\;\;\beta}\right)
	+ (1-\delta)\,i m_q^2 \bar{\psi}\gamma^\mu \left\langle {y^{\mu}\over y\cdot D}
	\right\rangle_{\!\!y}\psi
	\, ,
\end{eqnarray}
where $D$ is a covariant derivative operator, $y=(1,{\mathbf y})$ is a light like vector and 
$\langle \cdots \rangle$ is the average over
all possible directions, ${\hat y}$, of the loop momenta.
The HTL effective action is gauge invariant, nonlocal, and can generate all of the
HTL $n$-point functions~\cite{braaten,pisarski}, which are interrelated through 
Ward identities. The mass parameters $m_D$ and $m_q$ are the Debye screening and quark masses 
in a hot and dense medium, respectively,  which depend on the strong coupling $g$, temperature $T$, and 
the chemical potential $\mu$.  In the high temperature limit the leading-order expressions for $m_D$
and $m_q$ are
\begin{eqnarray}
m_D^2 &=& {g^2 \over 3} \bigg[\Big(N_c + {N_f \over 2}\Big) \, T^2 + \frac{3N_f}{2\pi^2}\mu^2 \bigg] \, , \\
m_q^2 &=& {g^2 \over 4} \frac{N_c^2-1}{4N_c} \bigg( T^2+\frac{\mu^2}{\pi^2}\bigg) \, .
\end{eqnarray}
We will not assume these expressions a priori, but instead treat $m_D$ and $m_q$ as free parameters to be fixed 
at the end of the calculation.  In order to make the calculation tractable we make expansions in $m_D$ and $m_q$  
in (\ref{L-HTL}) treating the masses as order $g$~\cite{andersen1,andersen2,andersen3}.
The $n^{\rm th}$ loop order in the HTLpt loop expansion is obtained by expanding the partition function through order $\delta^{n-1}$
and then taking $\delta \rightarrow 1$~\cite{andersen1,andersen2,andersen3,munshi,munshi1,munshi2,jiang}.
In this work, we will fix the parameters $m_D$ and $m_q$ by employing a variational prescription which
requires that the first derivative of the thermodynamic potential with respect to both $m_D$ and $m_q$ vanishes,
such that the free energy is minimized.  In the following, we generalize the NLO thermodynamic potential calculation from the case
of zero chemical potential~\cite{andersen2} to finite chemical potential.   
 
\section{Ingredients for the NLO Thermodynamic potential in HTLpt}
\label{ingredients}

The LO HTLpt thermodynamic potential, $\Omega_{\rm LO}$, for an $SU(N_c)$ gauge theory with $N_f$ massless quarks 
in the fundamental representation can be written as ~\cite{andersen1,andersen2}
\be
\Omega_{\rm LO}= d_A {\cal F}_{g}
+ d_F {\cal F}_q+\Delta_0{\cal E}_0\;,
\ee
where $d_F=N_f N_c$ and $d_A=N_c^2-1$ with $N_c$ is the number of
colors. ${\cal F}_q$ and ${\cal F}_g$ are the one loop contributions to quark and gluon free energies, respectively. 
The LO counterterm is the same as in the case of zero chemical potential~\cite{andersen1}
\be
 \Delta_0{\cal E}_0 = {d_A\over128\pi^2\epsilon}m_D^4\;.
\label{count0}
\ee

\begin{figure}[t]
\includegraphics[width=6.5cm]{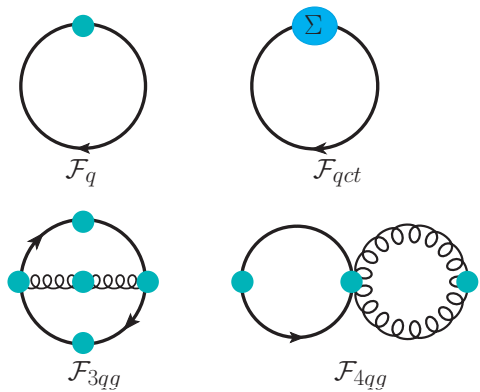}
\caption{Diagrams containing fermionic lines relevant for NLO thermodynamics potential in HTLpt with finite chemical potential.  
Shaded circles indicate HTL $n$-point functions.}
\label{diagramfig}
\end{figure}

At NLO one must consider the diagrams shown in Fig.~\ref{diagramfig}.
The resulting NLO HTLpt thermodynamic potential can be written in the following general form~\cite{andersen2} 
\be
\Omega_{\rm NLO}&=&\Omega_{\rm LO}+
d_A \left[{\cal F}_{3g}+{\cal F}_{4g}+{\cal F}_{gh}
+{\cal F}_{gct}
\right] 
+d_A s_F \left[{\cal F}_{3qg}+{\cal F}_{4qg}
\right] \nonumber \\
&&+ d_F {\cal F}_{qct}
+\Delta_1{\cal E}_0
+\Delta_1 m_D^2{\partial\over\partial m_D^2}
\Omega_{\rm LO}+\Delta_1 m_q^2{\partial\over\partial m_q^2}\Omega_{\rm LO}\;,
\label{OmegaNLO}
\ee
where $s_F=N_F/2$.
At NLO the terms that depend on the chemical potential 
are ${\cal F}_q$, ${\cal F}_{3qg}$, ${\cal F}_{4qg}$, ${\cal F}_{qct}$, 
$\Delta_1 m_q^2$, and $\Delta_1 m_D^2$ as displayed
in Fig.~\ref{diagramfig}. The other terms, e.g. ${\cal F}_g$, ${\cal F}_{3g}$, ${\cal F}_{4g}$, 
${\cal F}_{gh}$ and ${\cal F}_{gct}$ coming from gluon and ghost loops remain the same as the $\mu=0$ case~\cite{andersen2}. 
We also add that the vacuum energy counterterm, $\Delta_1{\cal E}_0$, remains the 
same as the $\mu=0$ case whereas the 
mass counterterms, $\Delta_1 m_D^2$ and $\Delta_1m_q^2$, have to be computed for $\mu\ne0$. These counterterms are 
of order $\delta$.  This completes a general description of 
contributions one needs to compute in order to determine NLO HTLpt thermodynamic potential 
at finite chemical potential.  We now proceed to the scalarization of the necessary diagrams.

\section{Scalarization of the fermionic diagrams}
\label{scalarint}

\noindent
The one-loop quark contribution coming from the first diagram in Fig.~\ref{diagramfig} can be written as 
\be
{\cal F}_q &=& -\suminto\log\det\left[P\!\!\!\!/-\Sigma(P)\right]\;
          = -2\suminto\log P^2-2\suminto
\log\left[{A_S^2-A_0^2\over P^2}\right]\;,
\label{qlead}
\ee
where 
\be
A_0(P)&=&iP_0-{m_q^2\over iP_0}{\cal T}_P\;,
\label{aodef}
\\ 
A_S(P)&=&|{\bf p}|+{m^2_q\over |{\bf p}|}\left[1-{\cal T}_P\right]\;,
\label{asdef}
\ee
and ${\cal T}_P$ is defined by the following integral~\cite{andersen2} 
\be
{\cal T}_P&=&\left\langle \frac{P_0^2}{P_0^2+p^2c^2}\right\rangle_c 
= {\omega(\epsilon)\over 2}\int\limits_{-1}^1 dc \, (1-c^2)^{-\epsilon}\frac{iP_0}{iP_0-|{\bf p}|c} \, , 
\label{def-tf}
\ee
with $w(\epsilon) = 2^{2\epsilon}\,\Gamma(2-2\epsilon)/\Gamma^2(1-\epsilon)$.
In three dimensions $\epsilon \rightarrow 0$ and (\ref{def-tf}) reduces to 
\be
{\cal T}_P&=&\frac{iP_0}{2|{\bf p}|}\log \frac{iP_0+|{\bf p}|}{iP_0-|{\bf p}|} \, ,
\label{def-tf1}
\ee 
with $P\equiv (P_0,{\bf p})$.  
In practice, one must use the general form and only take the
limit $\epsilon \rightarrow 0$ after regularization/renormalization.

The HTL quark counterterm at one-loop order can be rewritten from the second diagram in Fig.~\ref{diagramfig} as
\be
\label{ct1}
{\cal F}_{qct}=-4\suminto
{P^2+m^2_q\over A_S^2-A_0^2}\;.
\ee

The two-loop contributions coming from the third and fourth diagrams in Fig.~\ref{diagramfig} are given, respectively, by
\be
{\cal F}_{3qg}&=&{1\over2}g^2\sumint\mbox{Tr}\left[\Gamma^{\mu}(P,Q,R)S(Q)
\times \Gamma^{\nu}(P,Q,R)
S(R)\right]\Delta_{\mu\nu}(P) \, ,
\label{3qg}
\\  
{\cal F}_{4qg}&=&{1\over2}g^2\sumint\mbox{Tr}\left[\Gamma^{\mu\nu}(P,-P,Q,Q)S(Q)\right]
\Delta_{\mu\nu}(P)\;,
\label{4qg}
\ee
where $S$ is the quark propagator which is given by $S = (\gamma^\mu {\cal A}_\mu)^{-1}$ with
${\cal A}_\mu = (A_0(P),A_S(P)\hat{\bf p})$ and $\Delta^{\mu\nu}$ is the gluon propagator.   
The general covariant gauge $\Delta^{\mu\nu}$ can be expressed most conveniently in Minkowski space
\bqa\nonumber
\Delta^{\mu\nu}(p) &=&
\left[ - \Delta_T(p) g^{\mu \nu} + \Delta_X(p) n^\mu n^\nu \right]
- {n \!\cdot\! p \over p^2} \Delta_X(p)
        \left( p^\mu n^\nu  + n^\mu p^\nu \right)
+ \left[ \Delta_T(p) + {(n \!\cdot\! p)^2 \over p^2} \Delta_X(p)
        - {\xi \over p^2} \right] {p^\mu p^\nu \over p^2} \; , \nonumber \\
\label{gprop-TC}
\eqa
where $n^\mu$ is thermal rest frame four-vector and $\Delta_T$ and $\Delta_L$ are the transverse and longitudinal propagators
\bqa
\Delta_T(p)&=&{1 \over p^2-\Pi_T(p)}\;,
\label{Delta-T:M}
\\
\Delta_L(p)&=&{1 \over - n_p^2 p^2+\Pi_L(p)}\; ,
\label{Delta-L:M}
\eqa
with $n_p^{\mu} \;=\; n^{\mu} - (n_\mu p^\mu/p^2) p^{\mu}$.
It is convenient to introduce the following linear combination of transverse
and longitudinal propagators which turn out to make the calculations easier to
manage in practice
\bqa
\Delta_X(p)&=& \Delta_L(p) + \frac{1}{n_p^2}\Delta_T(p) \;.
\label{Delta-X:E}
\eqa
Also above $\Gamma^\mu$ and $\Gamma^{\mu\nu}$ are HTL-resummed $3$- and $4$-point functions. Many more details
concerning the HTL $n$-point functions including the general Coulomb gauge propagator etc. 
can be found in appendices of Refs.~\cite{andersen1,andersen2}.

In general covariant gauge, the sum of (\ref{3qg}) and (\ref{4qg})
reduces to 
\be
{\cal F}_{3qg+4qg}&=&{1\over2}g^2\sumint\Bigg\{
\Delta_X(P)\mbox{Tr}\left[
\Gamma^{00}S(Q)
\right]
-\Delta_T(P)\mbox{Tr}\left[
\Gamma^{\mu}S(Q)\Gamma^{\mu}S(R')
\right]
\nonumber \\
&& \hspace{3cm}
\label{coll}
+\Delta_X(P)
\mbox{Tr}\left[
\Gamma^{0}S(Q)\Gamma^{0}S(R')
\right]\Bigg\}\;,
\ee
where $\Delta_T$ is the transverse gluon propagator,
$\Delta_X$ is a combination of the longitudinal and transverse gluon propagators~\cite{andersen2}, 
and $R' = Q - P$.  After performing the traces 
of the $\gamma$-matrices one obtains~\cite{andersen2}
\be
{\cal F}_{3qg+4qg}&=&
- g^2\sumint{1\over A_S^2(Q)-A_0^2(Q)}
\Bigg\{
2(d-1)\Delta_T(P){\hat{\bf q}\!\cdot\!\hat{\bf r}A_S(Q)A_S(R)-A_0(Q)A_0(R)
\over A_S^2(R)-A_0^2(R)} \nonumber \\
&& -2\Delta_X(P)
{A_0(Q)A_0(R)+A_S(Q)A_S(R)\hat{\bf q}\!\cdot\!\hat{\bf r}
\over A_S^2(R)-A_0^2(R) } 
\nonumber \\
&&
-4m_q^2
\Delta_X(P)
\Bigg\langle
{A_0(Q)-A_s(Q)\hat{{\bf q}}\!\cdot\!\hat{{\bf y}}\over(P\!\cdot\!Y)^2
-(Q\!\cdot\!Y)^2}{1\over(Q\!\cdot\!Y)}\Bigg\rangle_{\!\!\bf \hat y} \nonumber \\
&&
+{8m_q^2\Delta_T(P)\over A_S^2(R)-A_0^2(R)}
\Bigg\langle
{(A_0(Q)-A_S(Q)\hat{\bf q}\!\cdot\!\hat{\bf y})
(A_0(R)-A_S(R)\hat{\bf r}\!\cdot\!\hat{\bf y})
\over(Q\!\cdot\!Y)(R\!\cdot\!Y)}
\bigg\rangle_{\!\!\bf \hat y}
\nonumber \\
&& 
+ {4 m_q^2\Delta_X(P)\over A_S^2(R)-A_0^2(R) }
\Bigg\langle
{2A_0(R)A_S(Q)\hat{\bf q}\!\cdot\!\hat{\bf y}-A_0(Q)A_0(R)-A_S(Q)A_S(R)
\hat{\bf q}\!\cdot\!\hat{\bf r}
\over(Q\!\cdot\!Y)(R\!\cdot\!Y)}
\Bigg\rangle_{\!\!\bf \hat y} \Bigg\} 
\nonumber
\\
&& 
+ O(g^2 m_q^4) \; ,
\label{2loop_si}
\ee
where $A_0$ and $A_S$ are defined in (\ref{aodef}) and (\ref{asdef}), respectively. We add that 
the exact evaluation of two-loop free energy could be performed numerically and would involve 
5-dimensional integrations; however, one would need to be able to identify all divergences and 
regulate the numerical integration appropriately.  Short of this, one can calculate the sum-integrals by expanding in
a power series in $m_D/T$, $m_q/T$, and $\mu/T$ in order to obtain semi-analytic expressions. 

\section{High temperature expansion}
\label{expand}

As discussed above, we make an expansion of two-loop free energies in a
power series of $m_D/T$ and $m_q/T$ to obtain a series which is nominally accurate to
order $g^5$.  
By ``nominally accurate'' we mean that we expand the scalar integrals treating $m_D$ and $m_q$
as ${\cal O}(g)$ keeping all terms which contribute through ${\cal O}(g^5)$; however, 
the resulting series is accurate to order $g^5$ in name
only.  At each order in HTLpt the result is an infinite series in $g$. Using the mass 
expansion we keep terms through order $g^5$ at all loop-orders of HTLpt in order to make
the calculation tractable.  At LO one obtains only the correct perturbative coefficients 
for the $g^0$ and $g^3$ terms when one expands in a strict power series in $g$.  At NLO 
one obtains the correct $g^0$, $g^2$, and $g^3$ coefficients and at NNLO one obtains
the correct $g^0$, $g^2$, $g^3$, $g^4$, and $g^5$ coefficients.  The resulting approximants
obtained when going from LO to NLO to NNLO are expected to show improved convergence since the loop
expansion is now explicitly expanded in terms of the relevant high-temperature degrees of
freedom (quark and gluon high-temperature quasiparticles).

In practice, the HTL $n$-point functions can have both hard and soft momenta scales on each leg.
At one-loop order the contributions can be classified ``hard'' or ``soft'' depending on whether
the loop momenta are order $T$ or $gT$, respectively; however, since the lowest fermionic
Matsubara mode corresponds to $P_0 = \pi T$, fermion loops are always hard.
The two-loop contributions to the thermodynamic potential 
can be grouped into hard-hard (hh), hard-soft (hs), and soft-soft (ss) contributions.
However, we note that one of the momenta contributing is always hard since it corresponds to
a fermionic loop and therefore there will be no two-loop soft-soft contribution. Below we calculate 
the various contributions to the sum-integrals presented in Sec.~\ref{scalarint}. 

\subsection{One-loop sum-integrals}
\label{1sis}

The one-loop sum-integrals (\ref{qlead}) and (\ref{ct1})  
correspond to the first two diagrams in Fig.~\ref{diagramfig}.  They represent
the leading-order quark contribution and order-$\delta$ HTL counterterm. We will
expand the sum-integrals through order $m_q^4$ taking $m_q$ to be of (leading) order $g$.
This gives a result which is nominally accurate (at one-loop) through order $g^5$.~\footnote{Of
course, this won't reproduce the full $g^5$ pQCD result in the limit $g\rightarrow0$.  
In order to reproduce all known coefficients through ${\cal O}(g^5)$, one would need to perform
a NNLO HTLpt calculation.}

\subsubsection{Hard Contribution}
\label{hard}

The hard contribution to the one-loop quark self-energy in (\ref{qlead}) can be expanded in
powers of $m_q^2$ as
\be
{\cal F}^{(h)}_q=
-2\suminto \log P^2-4m_q^2\suminto{1\over P^2}
+2m_q^4\suminto\!
\left[{2\over P^4}
-{1\over p^2P^2}+{2{\cal T}_P\over p^2P^2}
-{\left({\cal T}_P\right)^2\over p^2P_0^2}
\right] .
\label{1loop_exp}
\ee
Note that the function ${\cal T}_P$ does not appear in $m_q^2$ term.
The expressions for the sum-integrals in (\ref{1loop_exp}) are listed in Appendix~\ref{asis}.
Using those expressions, the hard contribution to the quark free energy becomes
\be
{\cal F}_q^{(h)} &=& -\frac{7\pi^2}{180}T^4\left(1+\frac{120}{7}\hat\mu^2 
            +\frac{240}{7}\hat\mu^4\right) + \left(\frac{\Lambda}
            {4\pi T}\right)^{2\epsilon} \frac{m_q^2 T^2}{6}\biggr[\left(
            1+12\hat\mu^2\right)
\nn &+& \left.
            \epsilon\left(2-2\ln2+2\frac{\zeta'(-1)}{\zeta(-1)}+ 24(\gamma+2\ln2)
            \ \hat\mu^2 - 28\ \zeta(3)\ \hat\mu^4 + {\cal O}\left(\hat\mu^6\right)
            \right)\right] 
\nn &+&
           \frac{m_q^4}{12\pi^2}(\pi^2 - 6) \, . 
\label{Quark1loop}
\ee
Expanding the HTL quark counterterm in~(\ref{ct1}) one can write
\be
{\cal F}^{(h)}_{\rm qct}=
4m_q^2\suminto{1\over P^2}
-4m_q^4\suminto\left[{2\over P^4}
-{1\over p^2P^2}
+{2\over p^2P^2}{\cal T}_P
-{1\over p^2P_0^2}\left({\cal T}_P\right)^2\right] ,
\label{1loop_ct}
\ee
where the expressions for various sum-integrals in (\ref{1loop_ct}) are listed in Appendix~\ref{asis}.
Using those expressions, the hard contribution to the HTL quark counterterm becomes
\be
{\cal F}_{qct}^{(h)} = -\frac{m_q^2 T^2}{6}\left(1+12\hat\mu^2\right)
-   \frac{m_q^4}{6\pi^2}(\pi^2 - 6)  \, .
\label{count}
\ee
We note that the first term in ~(\ref{count}) 
cancels the order-$\epsilon^0$ term in the coefficient of $m_q^2$ in (\ref{Quark1loop}).
There are no soft contributions either from the leading-order quark 
term in (\ref{qlead}) or from the HTL quark counterterm in (\ref{ct1}).

\subsection{Two-loop sum-integrals}
\label{2sis}

Since the two-loop sum-integrals given in~(\ref{coll}) 
contain an explicit factor of $g^2$, we only require an expansion
to order $m_q^2 m_D/T^3$ and $m_D^3/T^3$ in order to determine all terms contributing through order $g^5$.
We note that the soft scales are  given by $m_q$ and $m_D$ whereas the
hard scale is given by $T$, which leads to two different phase-space regions as discussed in Sec.~\ref{1sis}.
In the hard-hard region, all three momenta $P$, $Q$, and $R$ are hard whereas 
in the hard-soft region, two of the three momenta are hard and the other one is soft. 

\subsubsection{The hh contribution}

The self-energies for hard momenta are suppressed~\cite{braaten,pisarski,andersen2} by $m_D^2/T^2$
or $m_q^2/T^2$ relative to the propagators. For hard momenta, one just needs to expand in 
powers of gluon self-energies $\Pi_T$, $\Pi_L$, and quark self-energy $\Sigma$. So, the hard-hard 
contribution of $ {\cal F}_{3qg}$ 
and $ {\cal F}_{4qg}$ in (\ref{coll}) can be written as
\be
{\cal F}^{(hh)}_{3qg+4qg} &=&  (d-1)g^2\left[\ \sumint\frac{1}{P^2Q^2} - 2 \sumintb
\frac{1}{P^2Q^2}\right] + 2m_D^2g^2\ \sumintb\left[\frac{1}{p^2P^2Q^2}{\cal T}_P + \frac{1}
{P^4Q^2}- \frac{d-2}{d-1}\frac{1}{p^2P^2Q^2}\right]
\nn
 & + & m_D^2g^2\sumint\left[\frac{d+1}{d-1}\frac{1}{P^2Q^2r^2} - \frac{4d}{d-1}
\frac{q^2}{P^2Q^2r^4} - \frac{2d}{d-1}\frac{P\cdot Q}{P^2Q^2r^4}\right]{\cal T}_R
\nn
 &+& m_D^2g^2\sumint\left[\frac{3-d}{d-1}\frac{1}{P^2Q^2R^2} + \frac{2d}{d-1}
\frac{P\cdot Q}{P^2Q^2r^4} - \frac{d+2}{d-1}\frac{1}{P^2Q^2r^2}  + \frac{4d}{d-1}\frac{q^2}
{P^2Q^2r^4} - \frac{4}{d-1}\frac{q^2}{P^2Q^2r^2R^2}\right]
\nn
&+& 2m_q^2g^2(d-1) \sumint\left[\frac{1}{P^2Q_0^2Q^2} + \frac{p^2-r^2}
{q^2P^2Q_0^2R^2}\right]{\cal T}_Q + 2m_q^2g^2(d-1) \sumintb\left[\frac{2}{P^2Q^4} - \frac{1}
{P^2Q_0^2Q^2}{\cal T}_Q\right]
\nn
&+& 2m_q^2g^2(d-1) \sumint\left[\frac{d+3}{d-1}\frac{1}{P^2Q^2R^2} -\frac{2}{P^2Q^4}
- \frac{p^2-r^2}{q^2P^2Q^2R^2}\right] \, ,
\label{2loop_hh}
\ee
where the various sum-integrals are evaluated in Appendices~\ref{asis} and~\ref{bsis}. Using those sum-integral expressions, 
the hh contribution becomes
\be
{\cal F}^{(hh)}_{3qg+4qg} &=&  \frac{5\pi^2}{72}\frac{\alpha_s}{\pi}T^4\left[1 + \frac{72}{5}\ \hat\mu^2
+ \frac{144}{5}\ \hat\mu^4\right] 
\nn
 &-& \frac{1}{72}\frac{\alpha_s}{\pi}\left(\frac{\Lambda}
{4\pi T}\right)^{4\epsilon}\left[\frac{1+6(4-3\zeta(3))\ \hat\mu^2
-120(\zeta(3)-\zeta(5))\ \hat\mu^4 + {\cal O}
\left(\hat\mu^6\right)}{\epsilon} \right.\nn
&+& 1.3035 - 59.9055\ \hat\mu^2
- 75.4564\ \hat\mu^4 + {\cal O}\left(\hat\mu^6\right)\Big] m_D^2T^2
\nn&+& 
  \frac{1}{8}\frac{\alpha_s}{\pi}\left(\frac{\Lambda}{4\pi T}\right)^{4\epsilon}
\left[\frac{1-12\ \hat\mu^2}{\epsilon} + 8.9807 -152.793\ \hat\mu^2 +115.826\ \hat\mu^4 + {\cal O}\left
(\hat\mu^6\right)\right] m_q^2T^2 \, . 
\label{2loop_hh_f}
\ee

\subsubsection{The hs contribution}

Following Ref.~\cite{andersen2} one can extract the hard-soft contribution from~(\ref{coll}) as the momentum 
$P$ is soft whereas momenta $Q$ and $R$ are always hard. The function associated with 
the soft propagator $\Delta_T(0,{\bf p})$ or $\Delta_X(0,{\bf p})$
can be expanded in powers of the soft momentum ${\bf p}$. For
$\Delta_T(0,{\bf p})$, the resulting integrals over ${\bf p}$\
are not associated with any scale and they vanish in dimensional regularization.
The integration measure $\int_{\bf p}$ scales like $m_D^3$,
the soft propagator $\Delta_X(0,{\bf p})$ scales like $1/m_D^2$,
and every power of $p$ in the numerator scales like $m_D$.

The contributions that survive only through order $g^2 m_D^3 T$ 
and $m_q^2m_Dg^3T$ from  $ {\cal F}_{3qg}$ and $ {\cal F}_{4qg}$ in (\ref{coll}) 
are

\be \nonumber
{\cal F}_{3qg+4qg}^{(hs)}&=&g^2T\int_{\bf p}{1\over p^2+m^2_D}
\sumintoq\left[
{2\over Q^2}-{4q^2\over Q^4}\right]
+2m_D^2g^2T\int_{\bf p}{1\over p^2+m_D^2}
\sumintoq
\left[{1\over Q^4}
-{2(3+d)\over d}{q^2\over Q^6}+{8\over d}{q^4\over 
 Q^8}
\right]
\\ 
&&
-4m_q^2g^2T\int_{\bf p}{1\over p^2+m_D^2}
\sumintoq\left[{3\over Q^4}
-{4q^2\over Q^6} -{4\over Q^4} {\cal T}_Q
-{2\over Q^2}\left\langle {1\over(Q\!\cdot\!Y)^2} \right\rangle_{\!\!\bf \hat y}
\right]
\;.
\ee
Using the sum-integrals contained in Appendices \ref{asis} and \ref{bsis}, the hard-soft contribution 
becomes
\be
{\cal F}^{(hs)}_{3qg+4qg} = -\frac{1}{6}\alpha_s m_DT^3 (1+12\ \hat\mu^2) + \frac{\alpha_s}
{24\pi^2}\left[\frac{1}{\epsilon} + 1 + 2\gamma + 4\ln2 -14\zeta(3)\ \hat\mu^2 +62\zeta(5)
\ \hat\mu^4 + {\cal O}\left(\hat\mu^6\right)\right]
\nn
\times\left(\frac{\Lambda}{4\pi T}\right)
^{2\epsilon}\left(\frac{\Lambda}{2 m_D}\right)^{2\epsilon} m_D^3 T -\frac{\alpha_s}{2\pi^2}
m_q^2 m_DT \, . 
\label{2loop_hs_f}
\ee

\subsubsection{The ss contribution}

As discussed earlier in Sec.~\ref{expand} there is no soft-soft contribution from the diagrams in Fig.~\ref{diagramfig} since at least one of the loops is fermionic.

\subsection{Thermodynamic potential}
\label{subthpot}

Now we can obtain the HTLpt thermodynamic potential $\Omega(T,\mu,\alpha_s,m_D,m_q,\delta)$ through two-loop order
for which the contributions involving quark lines are computed here whereas the ghost and 
gluon contributions are computed in Ref.~\cite{andersen2}. We also follow the same prescription
as in Ref.~\cite{andersen2} to determine the mass parameter $m_D$ and $m_q$ from respective gap equations 
but with finite quark chemical potential, $\mu$.

\subsubsection{Leading order thermodynamic potential}

Using the expressions of ${\cal F}_q$ with finite quark chemical potential in (\ref{Quark1loop}) and 
${\cal F}_g$ from Ref.~\cite{andersen2}, the total contributions from the one-loop diagrams
including all terms through order $g^5$ becomes
\be
\Omega_{\rm one\;loop} &=& - d_A\frac{\pi^2T^4}{45}\left\{1+\frac{7}{4}\frac{d_F}{d_A}
\left(1+\frac{120}{7}\hat\mu^2+\frac{240}{7}\hat\mu^4\right) - \frac{15}{2}
\left[1+\epsilon\left(2+2\frac{\zeta'(-1)}{\zeta(-1)}+
2\ln{\frac{\hat\Lambda}{2}}\right)\right]\hat m_D^2 
\right.\nn&-&\left.
30\frac{d_F}{d_A}\left[\left(1+12\hat\mu^2\right)+\epsilon\left(2-2\ln2 + 
2\frac{\zeta'(-1)}{\zeta(-1)}+2\ln{\frac{\hat\Lambda}{2}} + 24(\gamma+2\ln2)\hat\mu^2
-28\zeta(3)\hat\mu^4+ {\cal O}\left(\hat\mu^6\right)\right)\right]\hat m_q^2
\right.\nn&+&\left.
\ 30\left(\frac{\Lambda}{2 m_D}\right)^{2\epsilon}\left[1+\frac{8}{3}\epsilon\right]
\hat m_D^3+\frac{45}{8}\left(\frac{1}{\epsilon}+2\ln{\frac{\hat\Lambda}{2}}-7+2\gamma
+ \frac{2\pi^2}{3}\right)\hat m_D^4 - 60\frac{d_F}{d_A}(\pi^2-6)\hat m_q^4\right\}
 \;,
\label{Omega-1loop}
\ee
where $\hat m_D$, $\hat m_q$, $\hat\Lambda$, and $\hat \mu$ are dimensionless variables:
\be
\hat m_D &=& {m_D \over 2 \pi T}  \;,
\\
\hat m_q &=& {m_q \over 2 \pi T}  \;,
\\
\hat \Lambda &=& {\Lambda \over 2 \pi T}  \;,
\\
\hat \mu &=& {\mu \over 2 \pi T}  \;. 
\ee
Adding the counterterm in (\ref{count0}),
we obtain the thermodynamic potential at leading order in the $\delta$-expansion:
\be
\Omega_{\rm LO}& =& - d_A\frac{\pi^2T^4}{45}\left\{1+\frac{7}{4}\frac{d_F}{d_A}\left(
1+\frac{120}{7}\hat\mu^2+\frac{240}{7}\hat\mu^4\right) - \frac{15}{2}\left[1+\epsilon
\left(2+2\frac{\zeta'(-1)}{\zeta(-1)}+2\ln{\frac{\hat\Lambda}{2}}\right)\right]\hat m_D^2 
\right.\nn&-&\left.
30\frac{d_F}{d_A}\left[\left(1+12\hat\mu^2\right)+\epsilon\left(2-2\ln2+2\frac{\zeta'(-1)}
{\zeta(-1)}+2\ln{\frac{\hat\Lambda}{2}} + 24(\gamma+2\ln2)\hat\mu^2-28\zeta(3)\hat\mu^4 +
 {\cal O}\left(\hat\mu^6\right)\right)\right]\hat m_q^2
\right.\nn&+&\left.
30\left(\frac{\Lambda}{2 m_D}\right)^{2\epsilon}\left[1+\frac{8}{3}\epsilon\right]
\ \hat m_D^3+\frac{45}{8}\left(2\ln{\frac{\hat\Lambda}{2}}-7+2\gamma+\frac{2\pi^2}{3}\right)
\hat m_D^4 - 60\frac{d_F}{d_A}(\pi^2-6)\hat m_q^4\right\}
 \;,
\label{Omega-LO}
\ee
where we have kept terms of ${\cal O}(\epsilon)$ since they will be needed for the
two-loop renormalization.

\subsubsection{Next-to-leading order thermodynamic potential}

The complete expression for the next-to-leading order correction to the thermodynamic potential
is the sum of the contributions from all two-loop diagrams, the quark and gluon counterterms,
and renormalization counterterms. Adding the contributions of the two-loop diagrams, ${\cal F}_{3qg+4qg}$, 
involving a quark line in (\ref{2loop_hh_f}) and (\ref{2loop_hs_f}) and the contributions of
${\cal F}_{3g+4g+gh}$ from Ref.~\cite{andersen2}, one obtains
\be
\Omega_{\rm two\;loop} &=& -d_A\frac{\pi^2T^4}{45}\frac{\alpha_s}{\pi}\left\{-\frac{5}{4}
                       \left[c_A+\frac{5}{2}s_F\left(1+\frac{72}{5}\ \hat\mu^2+\frac{144}{5}
                       \ \hat\mu^4\right)\right] + 15\left(c_A + s_F\left(1+12\ \hat\mu^2
                       \right)\right){\hat m_D}
\right.\nn&-& \left.
                       \frac{55}{8}\left[\left(c_A-\frac{4}{11}s_F\left[1+6(4-3\zeta(3))
                       \ \hat\mu^2 - 120(\zeta(3)-\zeta(5))\ \hat\mu^4 + {\cal O}\left(
                       \hat\mu^6\right)\right]\right)\left(\frac{1}{\epsilon}+4\ln{\frac
                       {\hat\Lambda}{2}}\right)
\right.\right.\nn &-& \left.\left.
                      s_F\left(0.4712 - 34.8761\ \hat\mu^2- 21.0214\ \hat\mu^4 +{\cal O}
                      \left(\hat\mu^6\right)\right)- c_A\left(\frac{72}{11}\ln{\hat m_D}
                     -1.96869\right) \right]{\hat m_D^2}
\right.\nn &-& \left. 
                     \frac{45}{2} s_F\left[\left(1-12\ \hat\mu^2\right)\left(\frac{1}
                     {\epsilon} + 4\ \ln{\frac{\hat\Lambda}{2}}\right) + 8.9807 -152.793
                     \ \hat\mu^2 +115.826\ \hat\mu^4 + {\cal O}\left(\hat\mu^6\right)\right]
                     {\hat m_q^2} 
\right.\nn &+& \left. 
                   180 s_F\ {\hat m_D}{\hat m_q^2} + \frac{165}{4}\left[\left(c_A-\frac{4}
                   {11}s_F\right)\left(\frac{1}{\epsilon} + 4\ \ln{\frac{\hat\Lambda}{2}} 
                  - 2\ln{\hat m_D}\right) + c_A\left(\frac{27}{11}+2\gamma\right)
\right.\right.\nn&-& \left.\left.
                  \frac{4}{11}s_F\left(1+2\gamma + 4\ln2 -14 \zeta(3)\ \hat\mu^2 + 62\zeta(5)
                  \ \hat\mu^4 + {\cal O}\left(\hat\mu^6\right)\right)\right] {\hat m_D^3}
                  \right\}\, ,
\label{Omega2loop}
\ee
where $c_A=N_c$ and $s_F=N_f/2$.
 
The HTL gluon counterterm is the same as obtained at zero chemical potential~\cite{andersen2} 
\be
\Omega_{\rm gct} = - d_A\frac{\pi^2 T^4}{45}\left[\frac{15}{2}\ {\hat m_D^2} - 45 \hat m_D^3
                    -\frac{45}{4}\left(\frac{1}{\epsilon} + 2\ln{\frac{\hat\Lambda}{2}} - 7 
                   + 2\gamma + \frac{2\pi^2}{3}\right) {\hat m_D^4}\right] ,
\label{OmegaGct}
\ee
The HTL quark counterterm as given by~(\ref{count}) is
\be
\Omega_{qct} = -d_F\frac{\pi^2 T^4}{45}\left[30(1 + 12\ \hat\mu^2)\ {\hat m_q^2} +
                     120(\pi^2-6)\ {\hat m_q^4}\right] \, .
\label{OmegaQct}
\ee

The ultraviolet divergences that remain after adding (\ref{Omega2loop}), (\ref{OmegaGct}), and (\ref{OmegaQct})
can be removed by renormalization of the vacuum energy density ${\cal E}_0$
and the HTL mass parameter $m_D$ and $m_q$.
The renormalization contributions~\cite{andersen2} at first order in $\delta$ are
\be
\Delta\Omega = \Delta_1{\cal E}_0 + \Delta_1 m_D^2\frac{\partial}{\partial m_D^2}\Omega_{LO}
+ \Delta_1 m_q^2\frac{\partial}{\partial m_q^2}\Omega_{LO} \, .
\label{del_tot}
\ee

The counterterm $\Delta_1{\cal E}_0$ at first order in $\delta$ will be same as the zero chemical
potential counterterm
\be
\Delta_1{\cal E}_0 = - {d_A\over 64\pi^2\epsilon} m_D^4 \, .
\label{del1e0}
\ee
The mass counterterms necessary at first order in $\delta$ are found to be
\be
\Delta_1 \hat m_D^2 = -\frac{\alpha_s}{3\pi\epsilon}\left[\frac{11}{4}c_A-s_F-s_F\left(1+6\hat m_D
                  \right)\left[(24-18\zeta(3))\hat\mu^2 + 120(\zeta(5)-\zeta(3))\hat\mu^4 + 
                  {\cal O}\left(\hat\mu^6\right)\right]\right]\ \hat m_D^2
\label{delmd}
\ee
and
\be
\Delta_1 \hat m_q^2 \;&=& -{\alpha_s\over 3 \pi \epsilon}\left[{9\over8}{d_A\over c_A}\right] 
\frac{1-12\ \hat\mu^2}{1+12\ \hat\mu^2}\ \hat m_q^2
\;.
\label{delmq}
\ee
Using the above counterterms, the complete contribution from the counterterms in (\ref{del_tot}) at first order
in $\delta$ at finite chemical potential becomes

\be
\Delta\Omega &=& - d_A\frac{\pi^2T^4}{45}\left\{\frac{45}{4\epsilon}\hat m_D^4+\frac{\alpha_s}
                 {\pi}\left[\frac{55}{8}\left(c_A-\frac{4}{11}s_F\left[1+(24-18\zeta(3))
                 \hat\mu^2 + 120(\zeta(5)-\zeta(3))\hat\mu^4 + {\cal O}\left(\hat\mu^6\right)
                 \right]\right)
\right.\right.\nn&&\left.\left.
                 \left(\frac{1}{\epsilon} + 2 + 2\frac{\zeta'(-1)}{\zeta(-1)}+2\ln{\frac{\hat
                 \Lambda}{2}}\right)\hat m_D^2 -\frac{165}{4}\left(c_A-\frac{4}{11}s_F\right)\left(\frac{1}{\epsilon}+2+
2\ln{\frac{\hat\Lambda}{2}} - 2\ln{\hat m_D}\right)\hat m_D^3
\right.\right.\nn
&-&\left.\left.
\frac{165}{4}\frac{4}{11}s_F\left[(24-18\zeta(3))\hat\mu^2 + 120(\zeta(5)-
\zeta(3))\hat\mu^4 + {\cal O}\left(\hat\mu^6\right)\right]\left(2\frac{\zeta'(-1)}{\zeta(-1)}
 +2\ln{\hat m_D} \right)\hat m_D^3
\right.\right.\nn
&+&\left.\left.
\frac{45}{2} s_F \frac{1-12\ \hat\mu^2}{1+12\ \hat\mu^2}\Bigg (\frac{1+12\ \hat\mu^2}{\epsilon}
+ 2 + 2\ln{\frac{\hat\Lambda}{2}} -2\ln 2
 +2{\zeta'(-1)\over\zeta(-1)} + 24(\gamma+2\ln2)\ \hat\mu^2 \right.\right.\nn 
&-& \left.\left.
28\zeta(3)\ \hat\mu^4 + {\cal O}
\left(\hat\mu^6\right)
 \Bigg ) \hat m_q^2
\right]\right\} .
\label{OmegaVMct}
\ee
Adding the contributions from the two-loop diagrams in~(\ref{Omega2loop}), the
HTL gluon and quark counterterms in~(\ref{OmegaGct}) and~(\ref{OmegaQct}), the
contribution from vacuum and mass renormalizations in~(\ref{OmegaVMct}), and
the leading-order thermodynamic potential in~(\ref{Omega-LO}) we
obtain the complete expression for the QCD thermodynamic potential 
at next-to-leading order in HTLpt:
\be
\Omega_{\rm NLO}&=&
           - d_A {\pi^2 T^4\over45} \Bigg\{ 
	   1 + {7\over4} {d_F \over d_A}\left(1+\frac{120}{7}
           \hat\mu^2+\frac{240}{7}\hat\mu^4\right) - 15 \hat m_D^3 
	  - {45\over4}\left(\log\hat{\Lambda\over2}-{7\over2}+\gamma+{\pi^2\over3}\right)\hat m_D^4	
\nn & + &
            60 {d_F \over d_A}\left(\pi^2-6\right)\hat m_q^4	
           + {\alpha_s\over\pi} \Bigg[ -{5\over4}\left(c_A + {5\over2}s_F\left(1+\frac{72}{5}
           \ \hat\mu^2+\frac{144}{5}\ \hat\mu^4\right)\right) 
	   + 15 \left(c_A+s_F(1+12\hat\mu^2)\right)\hat m_D
\nn & - & 
	 {55\over4}\left\{ c_A\left(\log{\hat\Lambda \over 2}- {36\over11}\log\hat m_D - 2.001\right)
		- {4\over11} s_F \left[\left(\log{\hat\Lambda \over 2}-2.337\right)\right.\right.
\nn & + & 
	\left.\left.  (24-18\zeta(3))\left(\log{\hat\Lambda \over 2} -15.662\right)\hat\mu^2
	+ 120\left(\zeta(5)-\zeta(3)\right)\left(\log{\hat\Lambda \over 2} -1.0811\right)\hat\mu^4 + 
         {\cal O}\left(\hat\mu^6\right)\right] \!\!\right\} \hat m_D^2
\nn & - & 
	45 \, s_F \left\{\log{\hat\Lambda\over 2} + 2.198  -44.953\hat\mu^2 - \left(288 \ln{\frac{\hat\Lambda}{2}} 
       +19.836\right)\hat\mu^4 + {\cal O}\left(\hat\mu^6\right)\right\} \hat m_q^2
\nn & + &
	 {165\over2}\left\{ c_A\left(\log{\hat\Lambda \over 2}+{5\over22}+\gamma\right)
	- {4\over11} s_F \left(\log{\hat\Lambda \over 2}-{1\over2}+\gamma+2\ln2 -7\zeta(3)\hat\mu^2+
         31\zeta(5)\hat\mu^4 + {\cal O}\left(\hat\mu^6\right) \right)\right\}\hat m_D^3
\nn & + &
         15 s_F \left(2\frac{\zeta'(-1)}{\zeta(-1)}
         +2\ln \hat m_D\right)\left[(24-18\zeta(3))\hat\mu^2 + 120(\zeta(5)-\zeta(3))\hat\mu^4 + 
           {\cal O}\left(\hat\mu^6\right)\right] \hat m_D^3
	+ 180\,s_F\hat m_D \hat m_q^2 \Bigg]
\Bigg\} .
\label{Omega-NLO}
\ee

\noindent
For convenience and comparison with lattice data \cite{fodor}, we define the pressure difference
\begin{eqnarray}
 \Delta P(T,\mu)=P(T,\mu)-P(T,0) \, .
\end{eqnarray}

\section{Pressure}
\label{nlopress}

In the previous section we have computed both LO and NLO thermodynamic potential in presence of quark chemical potential
and temperature. All other thermodynamic quantities can be calculated using standard thermodynamic relations. The pressure
is defined as 
\be
P = -{\Omega}(T,\mu,m_q,m_D) \, ,
\ee
where $m_D$ and $m_q$ are determined by requiring
\be
\frac{\partial \Omega_{\rm NLO}}{\partial \hat{m}_D} &=& 0 \, ,
\nonumber \\
\frac{\partial \Omega_{\rm NLO}}{\partial \hat{m}_q} &=& 0 \, .
\ee
This leads to the following two gap equations which will be solved numerically
\begin{eqnarray}
 45\hat m_D^2\left[1+\left(\ln\frac{\hat\Lambda}{2}-\frac{7}{2}+\gamma+
\frac{\pi^2}{3}\right)\hat m_D\right] &=&
\frac{\alpha_s}{\pi}\Bigg\{15(c_A+s_F(1+12\hat\mu^2))
-\frac{55}{2}\left[c_A\left(\ln\frac{\hat\Lambda}{2}-\frac{36}{11}\ln{\hat m_D}-3.637\right)
\right.
\nonumber\\
&&\left. \hspace{-3cm}
- \frac{4}{11}s_F\left\{\ln\frac{\hat\Lambda }{2}-2.333+(24-18\zeta(3))
\left(\ln\frac{\hat\Lambda }{2}-15.662\right)\hat\mu^2
\right.\right.
\nonumber\\
&&\left.\left.  \hspace{-3cm}
+ 120(\zeta(5)-\zeta(3))\left(\ln\frac{\hat\Lambda }{2}-1.0811\right)\hat\mu^4
\right\}\right]\hat m_D
+\frac{495}{2}\left[c_A\left(\ln\frac{\hat\Lambda }{2}
+\frac{5}{22}+\gamma\right)
\right.
\nonumber\\
&&\left.  \hspace{-3cm}
- \frac{4}{11}s_F\left\{\ln\frac{\hat\Lambda }{2}-\frac{1}{2}
+\gamma+2\ln2-7\zeta(3)\hat\mu^2+31\zeta(5)\hat\mu^4
\right.\right.
\nonumber\\
&&\left.\left.  \hspace{-3cm}
- \left(\frac{\zeta'(-1)}{\zeta(-1)}+\ln m_D+
\frac{1}{3}\right)\left((24-18\zeta(3))\hat\mu ^2+120(\zeta(5)-\zeta(3))\hat\mu^4\right)\right\}\right] m_D^2
+180 s_F\hat m_q^2\Bigg\} ,
\nonumber\\
\end{eqnarray}
and
\begin{eqnarray}
 \hat m_q^2=\frac{d_A}{8d_F\left(\pi ^2-6\right)}\frac{\alpha_s s_F}{\pi }\left[3\left(\ln\frac{\hat\Lambda }{2}
+2.198-44.953\ \hat\mu^2-\left(288\ln\frac{\hat\Lambda }{2}+19.836\right)\hat\mu^4\right)-12\hat m_D\right] ,
\end{eqnarray}

\begin{figure}
\subfigure{\includegraphics[width=8cm,height=8cm]{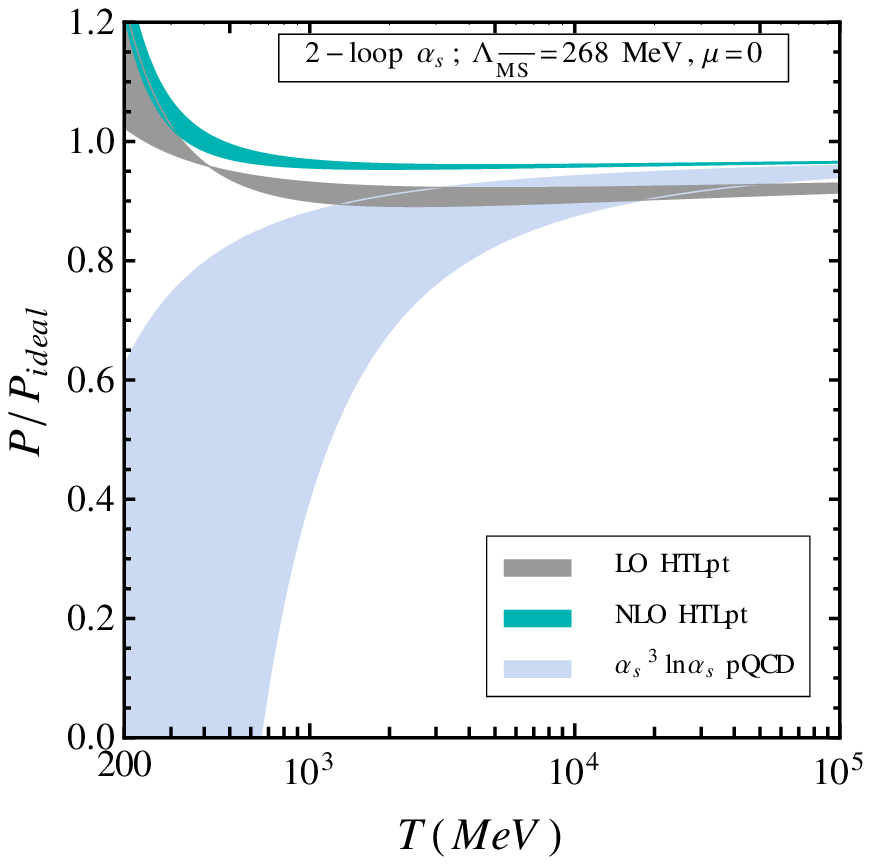}}
\subfigure{\includegraphics[width=8cm,height=8cm]{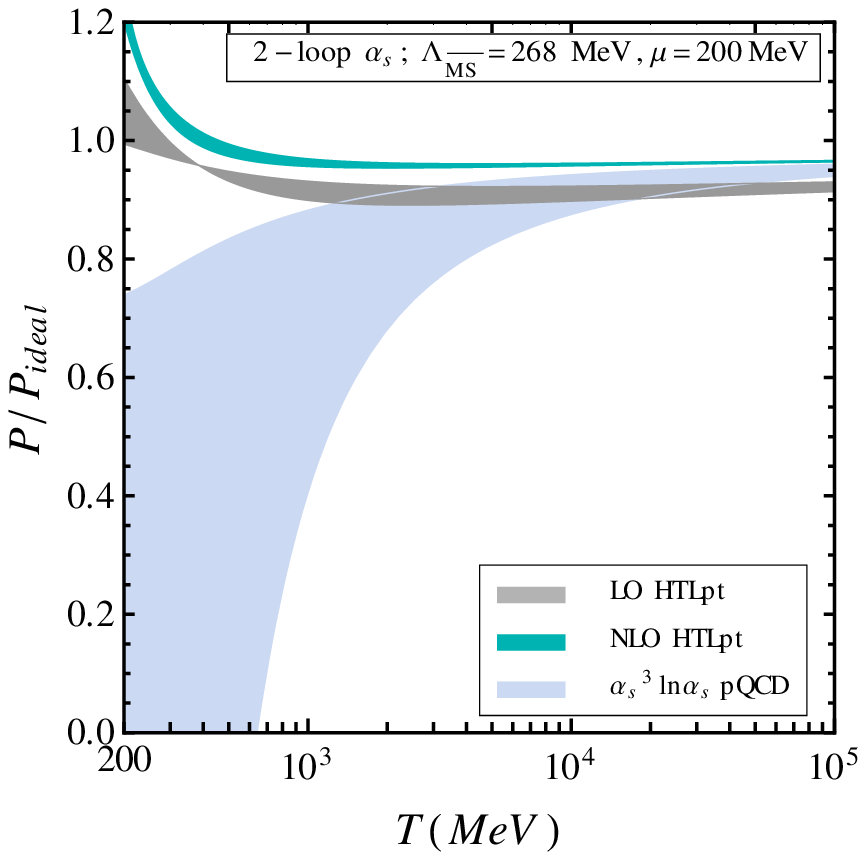}}
\caption{The NLO HTLpt pressure scaled with ideal gas pressure plotted along with four-loop
pQCD pressure~\cite{vuorinen} for two different values of chemical potential with $N_f=3$  and 2-loop running
coupling constant $\alpha_s$. The bands are obtained by varying the renormalisation scale by a factor of 
2 around its central value $\Lambda=2\pi \sqrt{T^2+\mu^2/\pi^2}$ \cite{vuorinen,scale}. 
We use $\Lambda_{\overline{\rm MS}}=290$ MeV based on recent lattice calculations~\cite{latt_lms} 
of the three-loop running of $\alpha_s$.}
\label{press_htl_nlo}
\end{figure}

\begin{figure}
\subfigure{\includegraphics[width=8cm,height=8cm]{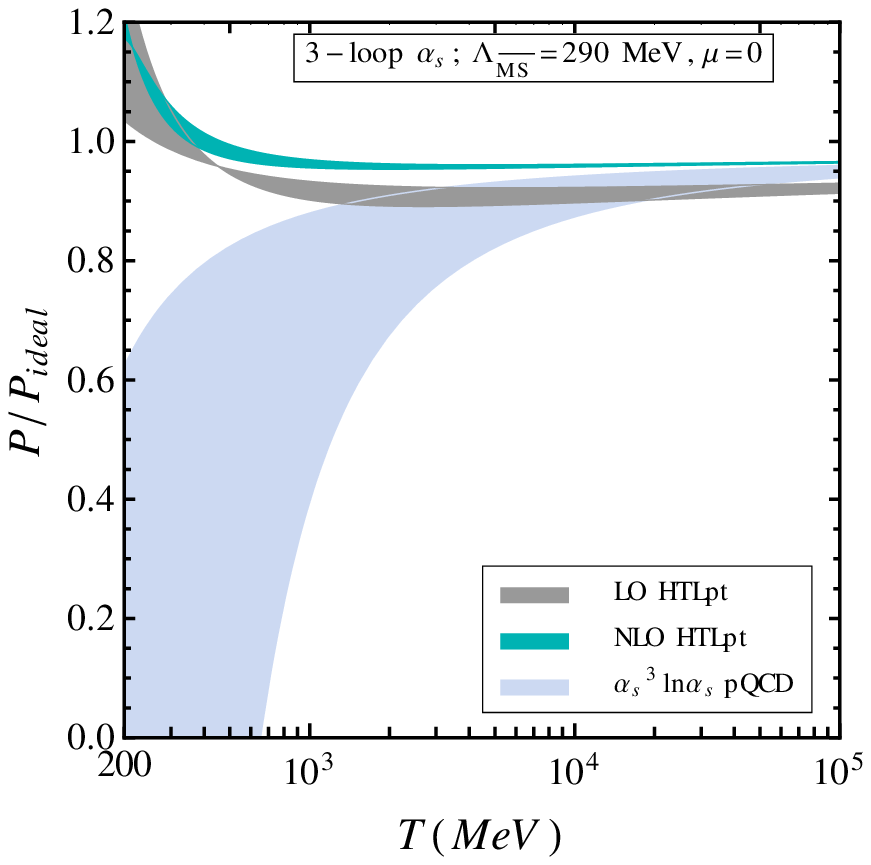}}
\subfigure{\includegraphics[width=8cm,height=8cm]{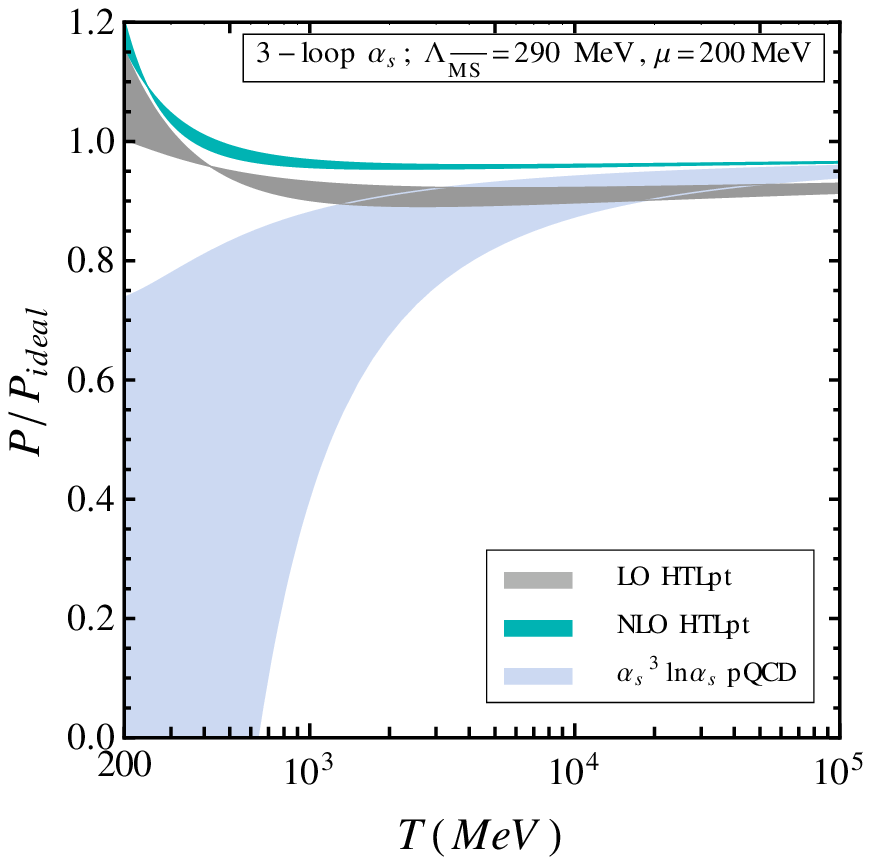}}
\caption{Same as Fig.~\ref{press_htl_nlo} but for 3-loop  $\alpha_s$.
}
\label{press_htl_nlo1}
\end{figure}

\begin{figure}
 \subfigure{
\includegraphics[width=8.5cm,height=8cm]{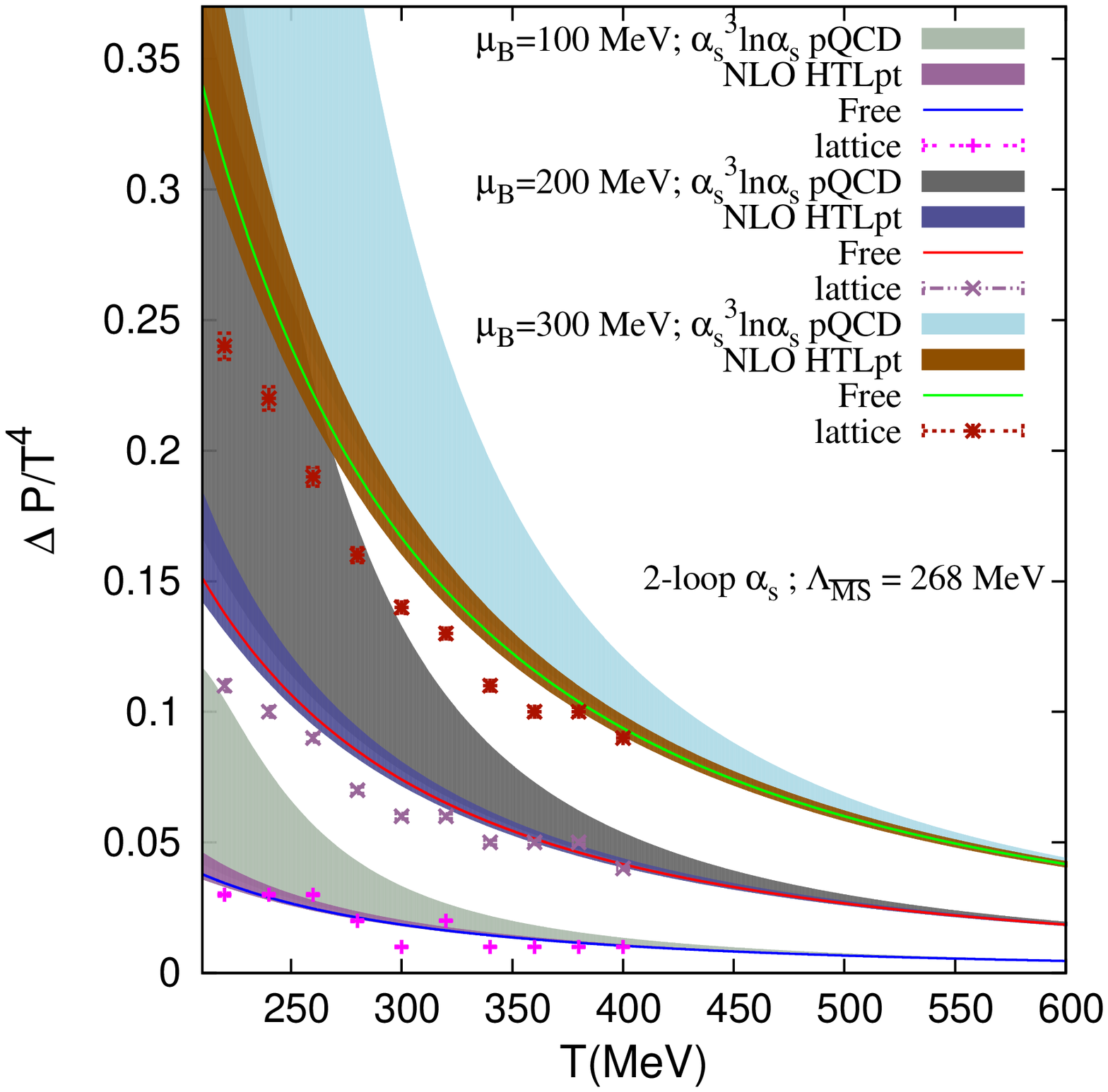}
}
\subfigure{
\includegraphics[width=8.5cm,height=8cm]{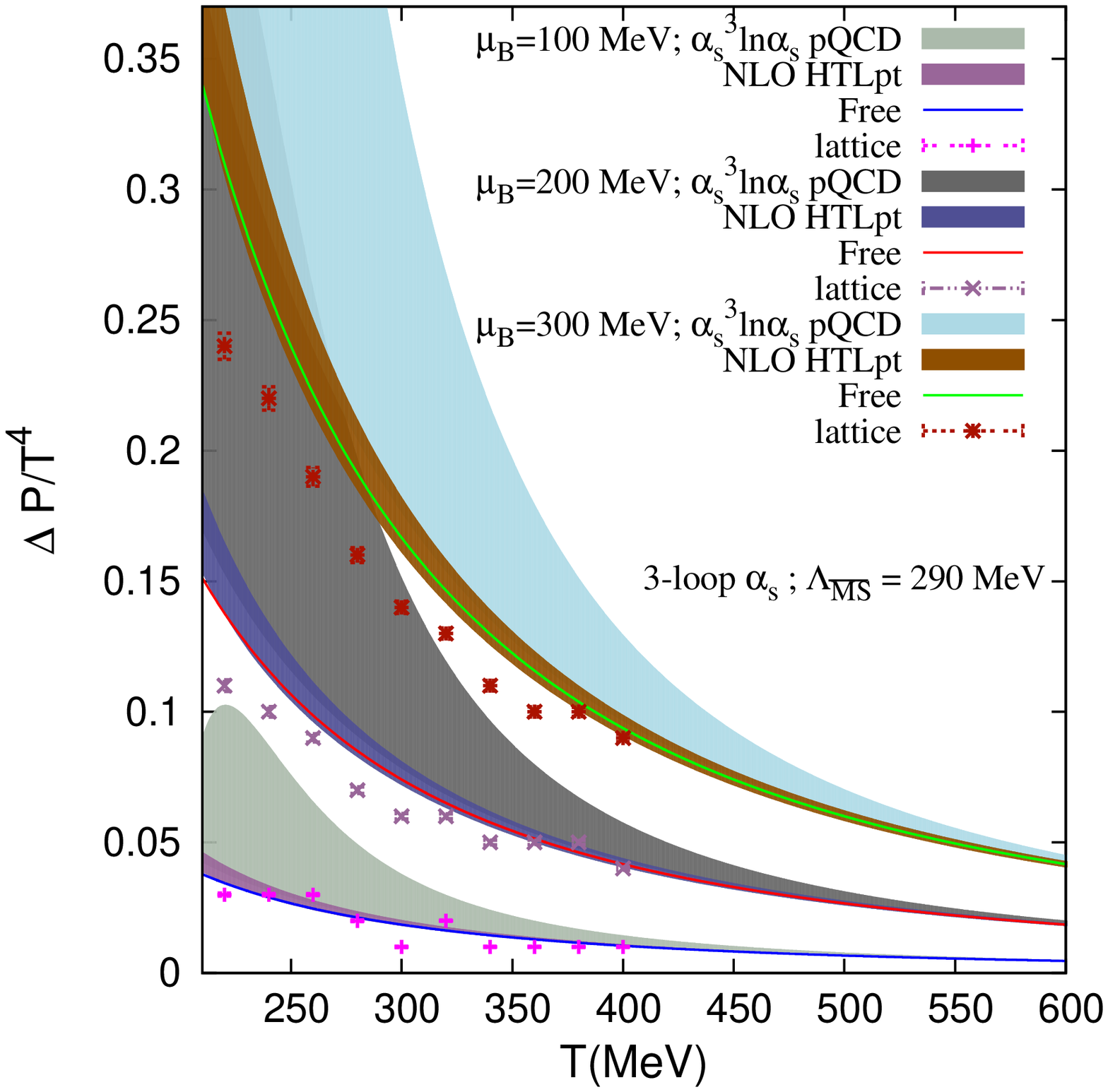}
}
\caption{(Left panel) $\Delta P$ for $N_f=3$ is plotted as a function of $T$ for two-loop
HTLpt result along with those of four-loop pQCD up to $\alpha_s^3\ln\alpha_s$ \cite{vuorinen} and 
lattice QCD~\cite{fodor} up to ${\cal O}\left(\mu^2\right)$ using 2-loop running coupling constant $\alpha_s$.  
(Right panel) Same as left panel but using 3-loop running coupling. 
In both cases three different values of $\mu$ are shown as specified in the legend. The bands in both HTLpt and pQCD are
obtained by varying the renormalisation scale by a factor of 2 around its  central value $\Lambda=2\pi\sqrt{T^2+\mu^2/\pi^2}$ \cite{vuorinen,scale}.
}
\label{diff_press}
\end{figure}

\noindent
where we note that we have discarded terms of ${\cal O}(\hat\mu^6)$ and higher above.  In Figs.~\ref{press_htl_nlo} and~\ref{press_htl_nlo1} we present a comparison of NLO HTLpt pressure with that of four-loop 
pQCD~\cite{vuorinen} as a function of the temperature for two and three loop running of $\alpha_s$.  
The only difference 
between Figs.~\ref{press_htl_nlo} and~\ref{press_htl_nlo1} is the choice of order of the running coupling used.  As can be seen from these figures, the dependence on the order of the running coupling is quite small. However, 
we note that in both figures even at extremely large
temperatures there is a sizable correction when going from LO to NLO.  This was already seen in the $T=0$
results of Ref.~\cite{andersen2} where it was found that due the logarithmic running of the coupling,
it was necessary to go to very large temperatures in order for the LO and NLO predictions to overlap.  This
is due to over-counting problems at LO which lead to an order-$g^2$ perturbative coefficient which is twice
as large as it should be \cite{andersen2,Haque:2013qta}.  This problem is corrected at NLO, but the end result is that there is a reasonably large correction ($\sim 5\%$) at the temperatures shown.

The NLO HTLpt result differs from 
the pQCD result through order $\alpha_s^3\ln\alpha_s$ at low temperatures.
A NNLO HTLpt calculation at finite $\mu$ would
agree better with pQCD $\alpha_s^3\ln\alpha_s$ as found in $\mu=0$ case~\cite{andersen3}. The  HTLpt result 
clearly indicates a modest improvement over pQCD in respect of convergence and sensitivity of the renormalisation 
scale. In Fig.~\ref{diff_press} the pressure difference, $\Delta P$,  is also compared with the same quantity computed using
pQCD~\cite{vuorinen} and lattice QCD~\cite{fodor}. Both LO and NLO HTLpt results are less sensitive to 
the choice of the renormalisation scale than the weak coupling results with the inclusion of successive orders of approximation. 
Comparison 
with available lattice QCD data~\cite{fodor} suggests that HTLpt and pQCD cannot accurately account for the lattice QCD results
below approximately $3\,T_c$; however, the results are in very good qualitative agreement with the lattice QCD results without
any fine tuning.

\section{Conclusions and Outlook}
\label{concl} 

In this paper we have generalized the zero chemical potential NLO HTLpt calculation of the QCD thermodynamic potential \cite{andersen2} 
to finite chemical potential. We have obtained (semi-)analytic expressions for the thermodynamic potential at both LO and NLO in HTLpt.
The results obtained are trustworthy at high temperatures and small chemical potential since we performed an expansion in the
ratio of the chemical potential over the temperature.

This calculation will be useful for the study of finite temperature and chemical potential QCD matter.  This is important
in view of the ongoing RHIC beam energy scan and proposed heavy-ion experiments at FAIR. Using the NLO HTLpt thermodynamic potential, 
we have obtained a variational solution for both mass parameters, $m_q$ and $m_D$, and we have used this to obtain the pressure 
at finite temperature and chemical potential. When compared with the weak coupling expansion 
of QCD, the HTLpt pressure helps somewhat with the problem of oscillation of successive approximations found in pQCD. Furthermore, the 
scale variation of the NLO HTLpt result for pressure is smaller than that obtained with the weak coupling result. 
The HTLpt pressure shows some deviations from the lattice data below 3 $T_c$ which suggests that the calculation should be extended to 
NNLO.   In addition, getting better agreement with pQCD at low temperature will require going to NNLO.
This is indeed a very challenging job which represents work in progress.

We also note that, based on the results obtained herein, one can straightforwardly compute quark susceptibilities.  
In a forthcoming paper we will compare the NLO HTLpt results for quark susceptibilities with lattice data and 
other theoretical models of QCD matter.  

\section*{Acknowledgements}
We thank S. Borsanyi and N. Su for useful discussions.  
M.S. was supported by NSF grant No.~PHY-1068765.


\appendix
\renewcommand{\theequation}{\thesection.\arabic{equation}}

\section{Sum-Integrals}
\label{asis}

In the imaginary-time (Euclidean time) formalism for the field theory of a hot and dense medium, 
the 4-momentum $P=(P_0,{\bf p})$ is Euclidean with $P^2=P_0^2+{\bf p}^2$. 
The Euclidean energy $P_0$ has discrete values:
$P_0=2n\pi T$ for bosons and $P_0=(2n+1)\pi T-i\mu$ for fermions,
where $n$ is an integer running from $-\infty$ to $\infty$, $\mu$ is the quark chemical potential,
and $T=1/\beta$ is the temperature of the medium. Loop diagrams usually then involve sums over $P_0$ and integrals 
over ${\bf p}$. In dimensional regularization, the integral over spatial momentum  is generalized
to $d = 3-2 \epsilon$ spatial dimensions.
We define the dimensionally regularized sum-integral as
\be
  \sumintob& \;\equiv\; &
  \left(\frac{e^\gamma\Lambda^2}{4\pi}\right)^\epsilon\;
  T\sum_{P_0=2n\pi T}\:\int {d^{3-2\epsilon}p \over (2 \pi)^{3-2\epsilon}}\;,\\ 
  \suminto& \;\equiv\; &
  \left(\frac{e^\gamma\Lambda^2}{4\pi}\right)^\epsilon\;
  T\sum_{P_0=(2n+1)\pi T-i\mu}\:\int {d^{3-2\epsilon}p \over (2 \pi)^{3-2\epsilon}}\;,
\label{sumint-def}
\ee
where $3-2\epsilon$ is the dimension of space, $\Lambda$ is an arbitrary
momentum scale, $P$ is the bosonic loop momentum, and $\{P\}$ is the fermionic loop momentum.
The factor $(e^\gamma/4\pi)^\epsilon$
is introduced so that, after minimal subtraction 
of the poles in $\epsilon$
due to ultraviolet divergences, $\Lambda$ coincides 
with the renormalization
scale of the $\overline{\rm MS}$ renormalization scheme.

We describe below the technique of contour integration~\cite{lebellac,kapusta} in the complex plane to evaluate 
the frequency sum over $P_0$.  Consider a meromorphic function $f(P_0)$ that originates from a loop diagram, then one can write 
\be
T\!\!\!\!\! \!\!\!\!\!\sum_{P_0=(2n+1)\pi T-i\mu}\!\!\!\!\! \!\!\!\!\! f(P_0) &=&  T \oint_{C} {{dP_0} \over {2\pi i }} 
\, f(P_0) \, \frac{\beta}{2} \tanh \frac{\beta(iP_0-\mu)}{2} 
=-\frac{T}{2\pi i} \times \frac{\beta}{2} \times (2\pi i) \sum \mbox{Residues} \, ,
\label{freeq_sum}
\ee
provided $f(P_0)$ is regular in  $\mbox{Re}(iP_0)=\mu$ line as shown in Fig.~\ref{contour}. Below we demonstrate two examples, a simpler one
involving only loop momentum  and a complicated one involving fourth power of loop momentum and the HTL angular function, which would 
be relevant for evaluating sum-integrals:

\begin{figure}[t]
\includegraphics[width=8cm,angle=90]{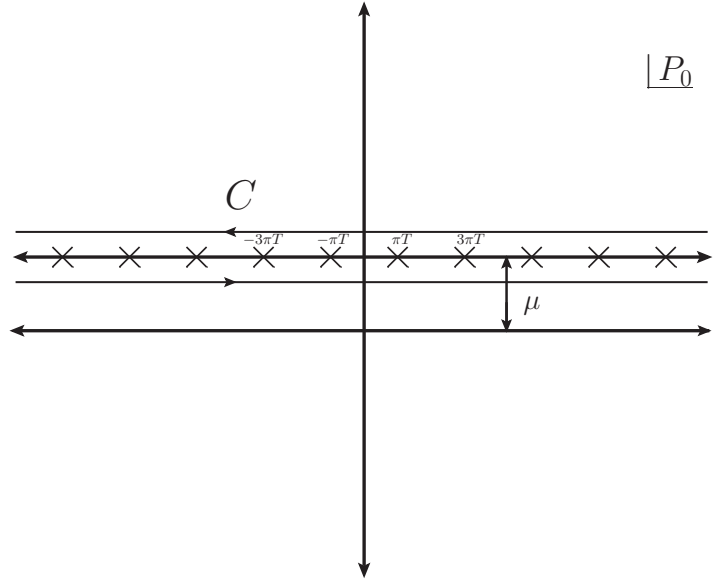}
\caption{The contour corresponding to (\ref{freeq_sum}) in complex $P_0$ plane. The crosses are
poles of the thermal weight factor which are shifted by an amount $\mu$ from the Re[$P_0$] axis.
}
\label{contour}
\end{figure}

\vspace{0.2in}

{\it (i) Simpler one:}

\be 
\suminto \frac{1}{P^2} &=&  \left(\frac{e^\gamma\Lambda^2}{4\pi}\right)^\epsilon\;
 \int {d^{3-2\epsilon}p \over (2 \pi)^{3-2\epsilon}}\;  T\sum_{P_0=(2n+1)\pi T-i\mu}\: \frac{1}{2p} 
\left [\frac{1}{iP_0+p}-\frac{1}{iP_0-p} \right ] \nonumber \\
&=& - \left(\frac{e^\gamma\Lambda^2}{4\pi}\right)^\epsilon\;
 \int {d^{3-2\epsilon}p \over (2 \pi)^{3-2\epsilon}}\; \left(\frac{n_F(p)}{2p}\right) ,
\ee
where
$n_F(p)=[e^{\beta(p-\mu)}+1]^{-1}+[e^{\beta(p+\mu)}+1]^{-1}=[n_F^-(p)+n_F^+(p)]$.

\vspace{0.2in}

{\it (ii) Involving HTL term:} 

\be
\suminto \frac{1}{P^4}{\cal T}_P &=&  \suminto \frac{1}{P^4}\left\langle\frac{P_0^2}{P_0^2+p^2c^2}
\right\rangle_c\nonumber\\
&=&\left\langle\frac{1}{1-c^2}\right\rangle_c \; \; \suminto \frac{1}{P^4} + \left\langle\frac{c^2}{1-c^2}\suminto 
\frac{1}{P^2(P_0^2+p^2c^2)}\right\rangle_c \nonumber\\
&=&\left\langle\frac{1}{1-c^2}\right\rangle_c \; \; \suminto \left(-\frac{1}{2p}\right)\frac{d}{dp}\frac{1}{P^2}
 + \left\langle\frac{c^2-c^{1+2\epsilon}}{(1-c^2)^2}\right\rangle_c\ \suminto 
\frac{1}{p^2P^2}\nonumber\\
&=& \frac{1}{2} \left(\frac{e^\gamma\Lambda^2}{4\pi}\right)^\epsilon\;
 \int {d^{3-2\epsilon}p \over (2 \pi)^{3-2\epsilon}}\; \frac{1}{p} \frac{d}{dp} \left [ \frac{n_F(p)}{2p}\right ]
-\left\langle\frac{c^2-c^{1+2\epsilon}}{(1-c^2)^2}\right\rangle_c 
\left(\frac{e^\gamma\Lambda^2}{4\pi}\right)^\epsilon\;
\int {d^{3-2\epsilon}p \over (2 \pi)^{3-2\epsilon}}\;\left [\frac{n_F(p)}{2p^3} \right ] .
\ee

After performing the frequency sum, one is left with dimensionally regularized spatial momentum integration,
which are also discussed in Appendix~\ref{bsis}. However, all other frequency sums can be evaluated in similar way as
discussed above. 

\subsection{Simple one loop sum-integrals}
\label{a1sis}

The specific fermionic one-loop sum-integrals needed are

\be 
\suminto \ \ln P^2 = \frac{7\pi^2}{360}T^4 \left(1 + \frac{120\ {\hat\mu^2}}{7}
+\frac{240\ {\hat\mu^4}}{7}\right) .
\ee
\be
\suminto \ \frac{1}{P^2}=-\frac{T^2}{24}\left(\frac{\Lambda}
{4\pi T}\right)^{2\epsilon}\left[1 + 12\ \hat\mu^2 + \epsilon\left(
2-2\ln2+2\frac{\zeta'(-1)}{\zeta(-1)} + 24(\gamma + 2\ln2)\ \hat\mu^2 
- 28\zeta(3)
\ \hat\mu^4 +{\cal O}\left(\hat\mu^6\right)\right)\right.\nn\left.
+\epsilon^2\left(4 + \frac{\pi^2}{4}-4\ln2 - 2\ln^22 + 4(1-\ln2)
\frac{\zeta'(-1)}{\zeta(-1)}+2\frac{\zeta''(-1)}{\zeta(-1)} + 94.5749\ \hat\mu^2 - 143.203\ \hat\mu^4 
+ {\cal O}\left(\hat\mu^6\right)\right )\right] . \;\;
\ee
\be
\suminto \ \frac{1}{P^4}=\frac{1}{(4\pi)^2}\left(\frac{\Lambda}
{4\pi T}\right)^{2\epsilon}\left[\frac{1}{\epsilon} + \left( 2\gamma + 4\ln2 
 -14\ \zeta(3)\ \hat\mu^2 + 62\ \zeta(5)\ \hat\mu^4 +
{\cal O}\left(\hat\mu^6\right) \right) 
\right.\nn\left.+ \epsilon\left( 4(2\gamma + \ln2)\ln 2 -4\gamma_1 + \frac{\pi^2}{4}
 - 71.6013\ \hat\mu^2 + 356.329\ \hat\mu^4 +{\cal O}\left(\hat\mu^6\right)\right)
\right] .
\ee
\be
\suminto \ \frac{p^2}{P^4}=-\frac{T^2}{16}\left(\frac{\Lambda}
{4\pi T}\right)^{2\epsilon}\left[1 +12\ \hat\mu^2 + \epsilon\left(\frac{4}{3} - 
2\ln2 +2\frac{\zeta'(-1)}{\zeta(-1)} + 8(3\gamma+6\ln2-1) \ \hat\mu^2
\right.\right.\nn\left.\left. 
- \ 28\ \zeta(3) \ \hat\mu^4 +{\cal O}
\left(\hat\mu^6\right)\right)\right] .
\ee
\be
\suminto \ \frac{p^2}{P^6}=\frac{1}{(4\pi)^2}\frac{3}{4}\left(\frac{\Lambda}
{4\pi T}\right)^{2\epsilon}\left[\frac{1}{\epsilon} + \left( 2\gamma 
-\frac{2}{3} + 4\ln2 
 -14\ \zeta(3)\ \hat\mu^2 + 62\ \zeta(5)\ \hat\mu^4 +
{\cal O}\left(\hat\mu^6\right) \right) 
\right] .
\ee
\be
\suminto \ \frac{p^4}{P^6}=-\frac{5T^2}{64}\left(\frac{\Lambda}
{4\pi T}\right)^{2\epsilon}\left[1 +12\ \hat\mu^2 + \epsilon\left(\frac{14}{15} - 
2\ln2 +2\frac{\zeta'(-1)}{\zeta(-1)} + 8\left(-\frac{8}{5}+3\gamma+6\ln2\right)\ \hat\mu^2 
\right.\right.\nn\left.\left.-\  
28\ \zeta(3)\ \hat\mu^4 +{\cal O}\left(\hat\mu^6\right)\right)\right] .
\ee
\be
\suminto \ \frac{p^4}{P^8}=\frac{1}{(4\pi)^2}\left(\frac{\Lambda}
{4\pi T}\right)^{2\epsilon}\frac{5}{8}\left[\frac{1}{\epsilon} + \left( 2\gamma 
-\frac{16}{15} + 4\ln2 
 -14\ \zeta(3)\ \hat\mu^2 + 62\ \zeta(5)\ \hat\mu^4 +
{\cal O}\left(\hat\mu^6\right) \right) 
\right] .
\ee
\be
\suminto \ \frac{1}{p^2P^2}=\frac{1}{(4\pi)^2}\left(\frac{\Lambda}
{4\pi T}\right)^{2\epsilon}2\left[\frac{1}{\epsilon} + \left(2+ 2\gamma 
+ 4\ln2  -14\ \zeta(3)\ \hat\mu^2 + 62\ \zeta(5)\ \hat\mu^4 +
{\cal O}\left(\hat\mu^6\right) \right)
\right.\nn\left. 
+\epsilon\left(4+8\ln2+4\ln^2 2+4\gamma +8\gamma\ln2+\frac{\pi^2}{4}-4\gamma_1
-105.259\ \hat\mu^2 + 484.908\ \hat\mu^4 + {\cal O}\left(\hat\mu^6\right)\right)
\right] .
\ee
\subsection{HTL one loop sum-integrals}
We also need some more difficult one-loop sum-integrals 
that involve the HTL function 
defined in (\ref{def-tf}).

The specific fermionic sum-integrals needed are

\be
\suminto \ \frac{1}{P^4}{\cal T}_P = \frac{1}{(4\pi)^2}\left(\frac{\Lambda}
{4\pi T}\right)^{2\epsilon}\frac{1}{2}\left[\frac{1}{\epsilon} + \left(1 + 2\gamma + 
4\ln2   -14\ \zeta(3)\ \hat\mu^2 + 62\ \zeta(5)\ \hat\mu^4 
+{\cal O}\left(\hat\mu^6\right)\right)\right] .
\ee
\be
\suminto \ \frac{1}{p^2P^2}{\cal T}_P = \frac{2}{(4\pi)^2}\left(\frac{\Lambda}
{4\pi T}\right)^{2\epsilon}  \left[\frac{\ln2}{\epsilon} + \left(\frac{\pi^2}{6} +
\ln2 \left( 2\gamma + 5\ln2   -14\ \zeta(3)\ \hat\mu^2 + 62\ \zeta(5)\ \hat\mu^4 
+{\cal O}\left(\hat\mu^6\right)\right)\right)
\right.\nn\left.
+ \ \epsilon\left(17.5137 - 85.398\ \hat\mu^2 + 383.629\ \hat\mu^4 +
{\cal O}\left(\hat\mu^6\right)\right)
\right]  . \;\;\;
\ee
\be
\suminto \ \frac{1}{P^2P_0^2}{\cal T}_P = \frac{1}{(4\pi)^2}\left(\frac{\Lambda}
{4\pi T}\right)^{2\epsilon}  \left[\frac{1}{\epsilon^2} + \frac{1}{\epsilon}2\left(
\gamma+2\ln2- 7\ \zeta(3)\ \hat\mu^2 + 31\ \zeta(5)\ \hat\mu^4 
 +{\cal O}\left(\hat\mu^6\right)\right)
\right.\nn\left.
+ \left(\frac{\pi^2}{4} + 4\ln^22 +8\gamma\ln2 -4\gamma_1 -71.6014\ \hat\mu^2 
+ 356.329\ \hat\mu^4 
+{\cal O}\left(\hat\mu^6\right)\right)\right] .
\ee
\be
\suminto \ \frac{1}{p^2P_0^2}\left({\cal T}_P\right)^2 = \frac{4}{(4\pi)^2}\left(\frac
{\Lambda}{4\pi T}\right)^{2\epsilon} \ln2 \left[\frac{1}{\epsilon} + \left(
 2\gamma + 5\ln2\right)  - 14\ \zeta(3)\ \hat\mu^2 + 62\ \zeta(5)\ \hat\mu^4 
+{\cal O}\left(\hat\mu^6\right)\right] .
\ee
\be
\suminto \ \frac{1}{P^2}\left\langle\frac{1}{(P\cdot Y)^2}\right\rangle_{\bf\hat y} =
- \frac{1}{(4\pi)^2}\left(\frac
{\Lambda}{4\pi T}\right)^{2\epsilon}  \left[\frac{1}{\epsilon} - 1+  2\gamma + 4\ln2
 - 14\ \zeta(3)\ \hat\mu^2 + 62\ \zeta(5)\ \hat\mu^4 
+{\cal O}\left(\hat\mu^6\right)\right] .
\ee
\subsection{Simple two loop sum-integrals}
\be
\sumint \ \frac{1}{P^2Q^2R^2}=\frac{T^2}{(4\pi)^2}\left(\frac{\Lambda}
{4\pi T}\right)^{4\epsilon}\left[\frac{\hat\mu^2}{\epsilon} + 2 (4\ln2+2\gamma+1)
\ \hat\mu^2-\frac{28}{3}\zeta(3)\ \hat\mu^4+{\cal O}\left(\hat\mu^6\right)\right] .
\label{2s1}
\ee
\be
\sumint \ \frac{1}{P^2Q^2r^2}=\frac{T^2}{(4\pi)^2}\left(\frac{\Lambda}
{4\pi T}\right)^{4\epsilon}\left(-\frac{1}{6}\right)\left[\frac{1}{\epsilon}
\left(1+12\hat\mu^2\right) + 4-2\ln2+ 4 \frac{\zeta'(-1)}{\zeta(-1)}
\right.\nn
+\left.48\left(1+\gamma+\ln2\right)\ \hat\mu^2
- 76\zeta(3)\ \hat\mu^4+{\cal O}\left(\hat\mu^6\right)\right] .
\label{2s2}
\ee
\be
\sumint \ \frac{p^2}{P^2Q^2r^4}=\frac{T^2}{(4\pi)^2}\left(\frac{\Lambda}
{4\pi T}\right)^{4\epsilon}\left(-\frac{1}{12}\right)\left[\frac{1}{\epsilon}
\left(1+12\hat\mu^2\right)+\left(\frac{11}{3}+2\gamma-2\ln2+2\frac{\zeta'(-1)}{\zeta(-1)}\right)
\right.\nn\left.
+ 4\left(7 + 12\gamma + 12\ln2 - 3\zeta(3)\right)\ \hat\mu^2
-4\left(27\zeta(3)-20\zeta(5)\right)\ \hat\mu^4+{\cal O}\left(\hat\mu^6\right)\right] .
\label{2s3}
\ee
\be
\sumint \ \frac{P\cdot Q}{P^2Q^2r^4}=\frac{T^2}{(4\pi)^2}\left(\frac{\Lambda}
{4\pi T}\right)^{4\epsilon}\left(-\frac{1}{36}\right)\left[1 - 6\gamma + 6 \frac{\zeta'(-1)}{\zeta(-1)}
+24\left\{2 +  3\zeta(3)\right\}
\ \hat\mu^2
\right.\nn\left.
+ \ 48(7\zeta(3)-10\zeta(5))\ \hat\mu^4+{\cal O}(\hat\mu^6)\right] .
\label{2s4}
\ee
\be
\sumint \ \frac{p^2}{r^2P^2Q^2R^2}=-\frac{T^2}{(4\pi)^2}\left(\frac{\Lambda}
{4\pi T}\right)^{4\epsilon}\frac{1}{72}\left[\frac{1}{\epsilon}
\left[1 -12(1-3\zeta(3))\ \hat\mu^2 + 240(\zeta(3)-\zeta(5))\ \hat\mu^4+{\cal O}
\left(\hat\mu^6\right)\right]
\right.\nn\left. - \left( 7.001 - 108.218\ \hat\mu^2
-304.034\ \hat\mu^4+{\cal O}(\hat\mu^6)\right)\right] .
\label{2s5}
\ee
\be
\sumint \ \frac{p^2}{q^2P^2Q^2R^2}&=&\frac{T^2}{(4\pi)^2}\left(\frac{\Lambda}
{4\pi T}\right)^{4\epsilon}\frac{5}{72}\left[\frac{1}{\epsilon}\left(1 - 
\frac{12}{5}\left(1+7\zeta(3)\right)\ 
\hat\mu^2 -\frac{24}{5}\left(14\zeta(3)-31\zeta(5)\right)\ \hat\mu^4+{\cal O}
\left(\hat\mu^6\right)\right)
\right.\nn
&&\left. \qquad\qquad \qquad \qquad \qquad\qquad +\left(9.5424 - 185.706\ \hat\mu^2
+916.268\ \hat\mu^4+{\cal O}(\hat\mu^6)\right)\right] .
\label{2s6}
\ee
\be
\sumint \ \frac{r^2}{q^2P^2Q^2R^2}&=& -\frac{T^2}{(4\pi)^2}\left(\frac{\Lambda}
{4\pi T}\right)^{4\epsilon}\frac{1}{18}\left[\frac{1}{\epsilon}\left(1 + 
3(-2+7\zeta(3))\ \hat\mu^2 + 6(14\zeta(3)-31\zeta(5))\ \hat\mu^4+{\cal O}(\hat\mu^6)\right)
\right.\nn
&& \qquad\qquad \qquad \qquad \qquad \qquad  \left.+\left(8.1428 + 96.9345 \ \hat\mu^2
- 974.609\ \hat\mu^4+{\cal O}(\hat\mu^6)\right)\right] .
\label{2s7}
\ee
The generalized two loop sum-integrals can be written from \cite{andersen2} as
\be
\sumint &&F(P) G(Q) H(R) =
\int\limits_{PQ} F(P) G(Q) H(R)
- \int\limits_{p_0,{\bf p}} \epsilon(p_0) n_F(|p_0|) \, 2 \, {\rm Im} F(-i p_0+ \varepsilon,{\bf p}) 
	\, {\rm Re} \int\limits_Q G(Q) H(R)\bigg|_{P_0 = -ip_0 + \varepsilon}
\nn && 
- \int\limits_{p_0,{\bf p}} \epsilon(p_0) n_F(|p_0|) \, 2 \, {\rm Im} G(-i p_0+ \varepsilon,{\bf p}) 
	\, {\rm Re} \int\limits_Q H(Q) F(R)\bigg|_{P_0 = -ip_0 + \varepsilon}
\\ && \nonumber
+ \int\limits_{p_0,{\bf p}} \epsilon(p_0) n_B(|p_0|) \, 2 \, {\rm Im} H(-i p_0+ \varepsilon,{\bf p}) 
	\, {\rm Re} \int\limits_Q F(Q) G(R)\bigg|_{P_0 = -ip_0 + \varepsilon}
\\ && \nonumber 
+ \int\limits_{p_0,{\bf p}} \epsilon(p_0) n_F(|p_0|) \, 2 \, {\rm Im} F(-i p_0+ \varepsilon,{\bf p}) \,
        \int\limits_{q_0,{\bf q}} \epsilon(q_0) n_F(|q_0|) \, 2 \, {\rm Im} G(-i q_0+ \varepsilon,{\bf q}) 
	\, {\rm Re} H(R)\bigg|_{R_0 = i (p_0 + q_0)+ \varepsilon}
\\ && \nonumber
- \int\limits_{p_0,{\bf p}} \epsilon(p_0) n_F(|p_0|) \, 2 \, {\rm Im} G(-i p_0+ \varepsilon,{\bf p}) \,
        \int\limits_{q_0,{\bf q}} \epsilon(q_0) n_B(|q_0|) \, 2 \, {\rm Im} H(-i q_0+ \varepsilon,{\bf q}) 
	\, {\rm Re} F(R)\bigg|_{R_0 = i (p_0 + q_0)+ \varepsilon}
\\ &&  
- \int\limits_{p_0,{\bf p}} \epsilon(p_0) n_B(|p_0|) \, 2 \, {\rm Im} H(-i p_0+ \varepsilon,{\bf p}) \,
        \int\limits_{q_0,{\bf q}} \epsilon(q_0) n_F(|q_0|) \, 2 \, {\rm Im} F(-i q_0+ \varepsilon,{\bf q}) 
	\, {\rm Re} G(R)\bigg|_{R_0 = i (p_0 + q_0)+ \varepsilon}
\;.
\label{int-2loop}
\ee
After applying Eq.~(\ref{int-2loop}) and using the delta function to calculate the $P_0$ and $Q_0$
integrations, the sum-integral (\ref{2s5}) reduces to 
\be
\sumint \ \frac{1}{P^2Q^2R^2} =\int\limits_{\bf pq}\frac{n_F^{-}(p)-n_F^{+}(p)}{2p}\frac{n_F^{-}(q)
-n_F^{+}(q)}{2q}\frac{2p\ q}{\Delta(p,q,r)} \, ,
\ee
where 
\be
n_F^{\pm}(p) = \frac{1}{e^{\beta(p\pm\mu)}+1} \quad \mbox{and}
\quad \Delta(p,q,r)=p^4+q^4+r^4 -2(p^2q^2+q^2r^2+p^2r^2) = -4p^2q^2(1-x^2) \, ,
\ee
and using the result of Eq.~(\ref{th-2s1}), we get sum-integral (\ref{2s1}) and agree with \cite{vuorinen}.


After applying Eq.~(\ref{int-2loop}), the sum-integral (\ref{2s2}) reduces to

\be
\sumint \ \frac{1}{P^2Q^2r^2} = -2 \int\limits_{\bf p}\frac{n_F(p)}{2 p}\int\limits_Q
\frac{1}{Q^2r^2} + \int\limits_{\bf pq}\frac{n_F(p)n_F(q)}{4pq}\frac{1}{r^2} \, ,
\ee
where $n_F(p)=n^-_F(p)+n^+_F(p)$ . Now using the result of 4-dimensional integrals from
\cite{andersen2} and applying Eq.~(\ref{th-f}) and Eq.~(\ref{th-2s2}), we can calculate sum-integral
Eq.~(\ref{2s2}). The sum-integrals (\ref{2s3}) can be calculated in same way:
 \be
\sumint \ \frac{p^2}{P^2Q^2r^4} = -2 \int\limits_{\bf p}\frac{n_F(p)}{2 p}\int\limits_Q
\frac{p^2}{Q^2r^4} + \int\limits_{\bf pq}\frac{n_F(p)n_F(q)}{4pq}\frac{p^2}{r^4}.
\ee
The sum-integral (\ref{2s4}) can be written as
\be
\sumint \ \frac{P\cdot Q}{P^2Q^2r^4}=\sumint \ \frac{P_0 Q_0}{P^2Q^2r^4}+
\frac{1}{2}\sumint \ \frac{1}{P^2Q^2r^2} - \sumint \ \frac{p^2}{P^2Q^2r^4}
\ee
Using Eq.~(\ref{int-2loop}) and after doing $P_0$ and $Q_0$ integrations, first  
sum-integral above reduces to
\be
\sumint \ \frac{P_0 Q_0}{P^2Q^2r^4} =  \int\limits_{\bf pq}\frac{n_F^{-}(p)-n_F^{+}(p)}{2\ p}
\ \frac{n_F^{-}(q)-n_F^{+}(q)}{2\ q}\ \frac{p\ q}{r^4} \, ,
\ee 
and the result is given in Eq.~(\ref{th-2s4}). 
The second term and third terms sum-integrals above are linear combinations of Eq.~(\ref{2s2})
and Eq.~(\ref{2s3}). Adding all of them, we get required sum-integral.


Similarly after  applying Eq.~(\ref{int-2loop}), the sum-integral (\ref{2s5}) reduces to 
\be
\sumint\ \frac{p^2}{r^2P^2Q^2R^2} = \left.\int\limits_{\bf p}\frac{n_B(p)}{p}\int\limits_Q
\frac{r^2}{p^2Q^2R^2}\right|_{P_0=-i p} - \left.\int\limits_{\bf p}\frac{n_F(p)}{2p}
\int\limits_Q\frac{1}{Q^2R^2}\left(\frac{q^2}{r^2}+\frac{p^2}{q^2}\right)\right|_{P_0=-i p} 
\nn
+ \int\limits_{\bf pq}\frac{n_F(p)n_F(q)}{4pq}\ \frac{q^2}{r^2}\ \frac{r^2-p^2-q^2}{
\Delta(p,q,r)} -  \int\limits_{\bf pq}\frac{n_F(p)n_B(q)}{4pq}\ \frac{p^2+r^2}{q^2}
\ \frac{r^2-p^2-q^2}{\Delta(p,q,r)} \, ,
\label{2s5_1}
\ee 
So
\be
\left\langle\frac{p^2+r^2}{q^2}\ \frac{r^2-p^2-q^2}{\Delta(p,q,r)}\right\rangle_{\hat p
\cdot\hat q} = \frac{1}{2 q^2\ \epsilon} \, ,
\ee
and
\be
\left\langle\frac{q^2}{r^2}\ \frac{r^2-p^2-q^2}{\Delta(p,q,r)}\right\rangle_{\hat p
\cdot\hat q} &=&\left\langle\frac{q^2}{\Delta(p,q,r)}\right\rangle_x - \left\langle
\frac{q^2(p^2+q^2)}{\Delta(p,q,r)}\right\rangle_x \, ,
\nn
&=&\frac{1-2\epsilon}{8\epsilon}\frac{1}{p^2} - \frac{1}{2\epsilon}\left\langle
\frac{q^2}{r^4}\right\rangle_x-\frac{1-2\epsilon}{8\epsilon}\frac{1}{p^2}
\nn 
&=& 
- \frac{1}{2\epsilon}\left\langle
\frac{q^2}{r^4}\right\rangle_x \, .
\ee
Using the above angular integration, Eq.~(\ref{2s5_1}) becomes
\be
\sumint\ \frac{p^2}{r^2P^2Q^2R^2} = \left.\int\limits_{\bf p}\frac{n_B(p)}{p}\int\limits_Q
\frac{r^2}{p^2Q^2R^2}\right|_{P_0=-i p} - \left.\int\limits_{\bf p}\frac{n_F(p)}{2p}
\int\limits_Q\frac{1}{Q^2R^2}\left(\frac{q^2}{r^2}+\frac{p^2}{q^2}\right)\right|_{P_0=-i p} 
\nn
-\frac{1}{2\epsilon} \ \int\limits_{\bf pq}\frac{n_F(p)n_F(q)}{4pq}\  
\frac{p^2}{r^4}  - \frac{1}{2 \epsilon}\ \int\limits_{\bf pq}\frac{n_F(p)n_B(q)}{4pq}\ 
 \frac{1}{ q^2} \, .
\label{2s5_2}
\ee 
Using the 4-dimensional integrals from \cite{andersen2} and Eqs.~(\ref{th-b}), ({\ref{th-f}}), 
({\ref{th-f1}}) and ({\ref{th-ff}}), 
we obtain the sum-integral (\ref{2s5}).

Similarly after  applying Eq.~(\ref{int-2loop}), the sum-integral (\ref{2s6}) reduces to 
\be
\sumint\ \frac{p^2}{q^2P^2Q^2R^2} = \left.\int\limits_{\bf p}\frac{n_B(p)}{p}\int\limits_Q
\frac{q^2}{Q^2r^2R^2}\right|_{P_0=-i p} - \left.\int\limits_{\bf p}\frac{n_F(p)}{2p}
\int\limits_Q\frac{1}{Q^2R^2}\left(\frac{p^2}{q^2}+\frac{q^2}{p^2}\right)\right|_{P_0=-i p} 
\nn
+ \int\limits_{\bf pq}\frac{n_F(p)n_F(q)}{4pq}\ \frac{p^2}{q^2}\ \frac{r^2-p^2-q^2}{
\Delta(p,q,r)} -  \int\limits_{\bf pq}\frac{n_F(p)n_B(q)}{4pq}\ \left(\frac{p^2}{r^2}
+\frac{r^2}{p^2}\right)
\ \frac{r^2-p^2-q^2}{\Delta(p,q,r)} \, .
\label{2s6_1}
\ee 
Now
\be
\left\langle\frac{p^2}{q^2}\ \frac{r^2-p^2-q^2}{\Delta(p,q,r)}\right\rangle_{\hat p
\cdot\hat q} = 0 \, ,
\ee
and
\be
\left\langle\left(\frac{p^2}{r^2}+\frac{r^2}{p^2}\right)\ \frac{r^2-p^2-q^2}{\Delta(p,q,r)}\right\rangle_{\hat p
\cdot\hat q} = 
\frac{1}{2\epsilon}\frac{1}{p^2}- \frac{1}{2\epsilon}\left\langle
\frac{p^2}{r^4}\right\rangle_x \, .
\ee
Using the above angular average, we find
\be
\sumint\ \frac{p^2}{q^2P^2Q^2R^2} &=& \left.\int\limits_{\bf p}\frac{n_B(p)}{p}\int\limits_Q
\frac{q^2}{Q^2r^2R^2}\right|_{P_0=-i p} - \left.\int\limits_{\bf p}\frac{n_F(p)}{2p}
\int\limits_Q\frac{1}{Q^2R^2}\left(\frac{p^2}{q^2}+\frac{q^2}{p^2}\right)\right|_{P_0=-i p} 
\nn
 &&-\frac{1}{2\epsilon}  \int\limits_{\bf pq}\frac{n_F(p)n_B(q)}{2pq}\ \frac{1}{p^2}
+ \frac{1}{2\epsilon}  \int\limits_{\bf pq}\frac{n_F(p)n_B(q)}{2pq}\ \frac{p^2}{r^4}
\label{2s6_1a}
\ee
Using the 4-dimensional integrals from \cite{andersen2} and Eqs.~(\ref{th-b}), ({\ref{th-f}}), ({\ref{th-f1}}) and ({\ref{th-fb}}), 
we obtain the sum-integral (\ref{2s6}).

Similarly after  applying Eq.~(\ref{int-2loop}), the sum-integral (\ref{2s7}) reduces to 
\be
\sumint\ \frac{r^2}{p^2P^2Q^2R^2} = \left.\int\limits_{\bf p}\frac{n_B(p)}{p}\int\limits_Q
\frac{p^2}{Q^2r^2R^2}\right|_{P_0=-i p} - \left.\int\limits_{\bf p}\frac{n_F(p)}{2p}
\int\limits_Q\frac{1}{Q^2R^2}\left(\frac{r^2}{p^2}+\frac{r^2}{q^2}\right)\right|_{P_0=-i p} 
\nn
+ \int\limits_{\bf pq}\frac{n_F(p)n_F(q)}{4pq}\ \frac{r^2}{p^2}\ \frac{r^2-p^2-q^2}{
\Delta(p,q,r)} -  \int\limits_{\bf pq}\frac{n_F(p)n_B(q)}{4pq}\ \left(\frac{q^2}{r^2}
+\frac{q^2}{p^2}\right)
\ \frac{r^2-p^2-q^2}{\Delta(p,q,r)} \, .
\label{2s7_1}
\ee 
Now
\be
\left\langle\frac{r^2}{p^2}\ \frac{r^2-p^2-q^2}{\Delta(p,q,r)}\right\rangle_{\hat p
\cdot\hat q} = \frac{1}{2p^2\epsilon}\, ,
\ee
and
\be
\left\langle\left(\frac{q^2}{r^2}+\frac{q^2}{p^2}\right)\ \frac{r^2-p^2-q^2}{\Delta(p,q,r)}\right\rangle_{\hat p
\cdot\hat q} = - \frac{1}{2\epsilon}\left\langle
\frac{q^2}{r^4}\right\rangle_x \, .
\ee
Using the above angular average, we have
\be
\sumint\ \frac{p^2}{q^2P^2Q^2R^2} &=& \left.\int\limits_{\bf p}\frac{n_B(p)}{p}\int\limits_Q
\frac{q^2}{Q^2r^2R^2}\right|_{P_0=-i p} - \left.\int\limits_{\bf p}\frac{n_F(p)}{2p}
\int\limits_Q\frac{1}{Q^2R^2}\left(\frac{p^2}{q^2}+\frac{q^2}{p^2}\right)\right|_{P_0=-i p} 
\nn
 &&+\frac{1}{2\epsilon}  \int\limits_{\bf pq}\frac{n_F(p)n_B(q)}{2pq}\ \frac{1}{p^2}
+ \frac{1}{2\epsilon}  \int\limits_{\bf pq}\frac{n_F(p)n_B(q)}{2pq}\ \frac{q^2}{r^4} \, .
\label{2s7_2}
\ee
Using the 4-dimensional integrals from \cite{andersen2} and Eqs.~(\ref{th-b}), 
({\ref{th-f}}), ({\ref{th-f1}}) and ({\ref{th-bf}}), 
we obtain the sum-integral (\ref{2s6}).


\subsection{HTL two loop sum-integrals}
\be
\sumint \ \frac{1}{P^2Q^2r^2}{\cal T}_R=\frac{T^2}{(4\pi)^2}\left(\frac{\Lambda}
{4\pi T}\right)^{4\epsilon}\left(-\frac{1}{48}\right)\left[\frac{1}{\epsilon^2}
+\left(2+12(1+8\ \hat\mu^2)\ln2 + 4\frac{\zeta'(-1)}{\zeta(-1)}\right)\frac{1}{\epsilon}
\right.\nn\left.
+\left(136.3618 + 460.23 \ \hat\mu^2 - 273.046\ \hat\mu^4+{\cal O}\left(\hat\mu^6\right)\right)\right] .
\ee

\be
\sumint \ \frac{p^2}{P^2Q^2r^4}{\cal T}_R=\frac{T^2}{(4\pi)^2}\left(\frac{\Lambda}
{4\pi T}\right)^{4\epsilon}\left(-\frac{1}{576}\right)\left[\frac{1}{\epsilon^2}
+\left(\frac{26}{3}+4(13 + 144\ \hat\mu^2)\ln2 + 4\frac{\zeta'(-1)}{\zeta(-1)}
\right)\frac{1}{\epsilon}
\right.\nn\left.
+\left(446.397 + 2717.86\ \hat\mu^2 - 1735.61\ \hat\mu^4+{\cal O}(\hat\mu^6)\right)\right] .
\ee

\be
\sumint \ \frac{P\cdot Q}{P^2Q^2r^4}{\cal T}_R=\frac{T^2}{(4\pi)^2}\left(\frac{\Lambda}
{4\pi T}\right)^{4\epsilon}\left(-\frac{1}{96}\right)\left[\frac{1}{\epsilon^2}
+\left(4\ln2 + 4\frac{\zeta'(-1)}{\zeta'(-1)}\right)\frac{1}{\epsilon}
+\left(69.1737 + 118.244\ \hat\mu^2
\right.\right. 
\nn
\left.\left.
+ 136.688\ \hat\mu^4+{\cal O}\left(\hat\mu^6\right)\right)\right] .
\ee

\be
\sumint \ \frac{r^2-p^2}{P^2q^2Q_0^2R^2}{\cal T}_Q &=&-\frac{T^2}{(4\pi)^2}\left(
\frac{\Lambda}{4\pi T}\right)^{4\epsilon}\frac{1}{8}\left[\frac{1}{\epsilon^2}\left(1+4
\ \hat\mu^2\right)+\frac{1}{\epsilon}\left(2+2\gamma+\frac{10}{3}\ln2+2\frac{\zeta'(-1)}{\zeta(-1)}
\right.\right. 
\nn
 &+&\left.\left. 
2\ (8\gamma + 16\ln2 -7\zeta(3)) \ \hat\mu^2-\frac{2}{3}\left(98\zeta(3)-93\zeta(5)\right)
\ \hat\mu^4 + {\cal O}\left(\hat\mu^6\right)\right)
\right.\nn
&+&\left.
\left(46.8757 - 41.1192  \ \hat\mu^2 + 
64.0841 \ \hat\mu^4+{\cal O}\left(\hat\mu^6\right)\right)\right] .
\ee

\section{Integrals}
\label{bsis}

\subsection{Three dimensional integrals}

We require one integral that does not involve the 
Bose-Einstein distribution function.
The momentum scale in these integrals is set by the mass
$m=m_D$.
The one-loop integral is
\be
\int_{\bf p} {1 \over p^2+m^2} & = & 
- {m\over 4\pi} \left( {\Lambda \over 2 m} \right)^{2 \epsilon}
\left[1 + 2 \epsilon  \right] \,.
\label{bi3}
\ee

\subsection{Thermal Integrals}
\be
\frac{{\Lambda}^{2\epsilon}}{(4\pi)^2}\int\limits_{\bf p}\frac{n_B(p)}{p}p^{-2\epsilon}
&=&
    \frac{T^2}{(4\pi)^2}\left(\frac{\Lambda}{4\pi T}\right)^{4\epsilon}\left(\frac{1}{12}
     \right)\left\{1 + \epsilon\left[2-2\ln2+ 4 \frac{\zeta'(-1)}{\zeta(-1)}\right] 
\right.\nn &+& \left.
     2\epsilon^2\ \left[\frac{7\pi^2}{8} - 2 + \ln^2 2 -2 \ln2 + 4(1+\ln2)\left(1+ \frac
     {\zeta'(-1)}{\zeta(-1)}\right) + 4 \frac{\zeta''(-1)}{\zeta(-1)}\right]\right\}.
\label{th-b}
\ee
\be
\frac{{\Lambda}^{2\epsilon}}{(4\pi)^2}\int\limits_{\bf p}\frac{n_F(p)}{2p}p^{-2\epsilon}
&=&
    \frac{T^2}{(4\pi)^2}\left(\frac{\Lambda}{4\pi T}\right)^{4\epsilon}\left(\frac{1}{24}
     \right)\Big[\left(1+12\hat\mu^2\right)+\epsilon\Big\{2-2\ln2+ 4 \frac{\zeta'(-1)}{\zeta(-1)}
\nn &+& 
    24\left(2\gamma+5\ln2-1\right)\ \hat\mu^2 -56\zeta(3)\ \hat\mu^4+{\cal O}\left(\hat\mu^6\right)\Big\}\Big].
\label{th-f}
\ee
\be
\frac{{\Lambda}^{2\epsilon}}{(4\pi)^2}\int\limits_{\bf p}\frac{n_F(p)}{2p}
\frac{1}{p^2}p^{-2\epsilon}
&=&
    -\frac{T^2}{(4\pi)^2}\left(\frac{\Lambda}{4\pi T}\right)^{4\epsilon}\left[\frac{1}
    {\epsilon} + 2 + 2\gamma+10\ln2 -28\zeta(3)\ \hat\mu^2 + 124\zeta(5)\ \hat\mu^4 
    + {\cal O}\left(\hat\mu^6\right)\right].
\label{th-f1}
\ee
\be
\int\limits_{\bf pq}\frac{n_F(p)n_F(q)}{4pq}\frac{1}{r^2} = 
           \frac{T^2}{(4\pi)^2}\left[\frac{1}{3}(1-\ln2) + 4(2\ln2-1)\hat\mu^2+\frac{10}{3}
           \zeta(3)\ \hat\mu^4+{\cal O}\left(\hat\mu^6\right)\right].
\label{th-2s2}
\ee
\be
\int\limits_{\bf pq}\frac{n_F(p)n_F(q)}{4pq}\frac{p^2}{r^4} = 
             \frac{T^2}{(4\pi)^2}\left(-\frac{1}{36}\right)\left[\Bigg(5 + 6\gamma + 
             6\ln2 - 6 \frac{\zeta'(-1)}{\zeta(-1)} - 12(-13 + 12 \ln2 + 3 \zeta(3))\hat\mu^2
\right.\nn\left.
             +12 \left(-13 \zeta(3) + 20 \zeta(5)\right)\ \hat\mu^4+{\cal O}\left(\hat\mu^6
             \right)\Bigg) + \epsilon\left(3.0747 + 31.2624 \ \hat\mu^2 +262.387 \ \hat\mu^4
             + {\cal O}\left(\hat\mu^6\right) \right)\right].
\label{th-ff}
\ee
\be
\int\limits_{\bf pq}\frac{n_B(p)n_F(q)}{2pq}\frac{p^2}{r^4}=\frac{T^2}{(4\pi)^2}\left(
-\frac{1}{36}\right)\left[\Bigg\{7 - 6\gamma -18\ln2 + 6 \frac{\zeta'(-1)}{\zeta(-1)}+ 6 (-22 + 21 \zeta(3))
\ \hat\mu^2\right.\nn\left. + 
6\left(126 \zeta(3) - 155 \zeta(5)\right)\ \hat\mu^4+{\cal O}\left(\hat\mu^6\right)\Bigg\} 
+ \epsilon\left (29.5113 +
158.176 \ \hat\mu^2 - 557.189\ \hat\mu^4 + {\cal O}\left(\hat\mu^6\right)\right)\right].
\label{th-fb}
\ee
\be
\int\limits_{\bf pq}\frac{n_B(p)n_F(q)}{2pq}\frac{q^2}{r^4}=\frac{T^2}{(4\pi)^2}\left(
\frac{1}{18}\right)\left[\Bigg \{ 1 -
6\gamma - 12\ln2 + 6 \frac{\zeta'(-1)}{\zeta(-1)} + 12 \hat\mu^2
-6 \left(28 \zeta(3) 
-31 \zeta(5)\right)\ \hat\mu^4 
\right.\nn
+ {\cal O}\left(\hat\mu^6\right) \Bigg \}
+ \epsilon\left(31.0735 + 222.294\ \hat\mu^2 -\ 416.474\ \hat\mu^4 + {\cal O}\left(
\hat\mu^6\right)\right)\Biggr].
\label{th-bf}
\ee
 \be
\int\limits_{\bf pq}\frac{n_F^{-}(p)-n_F^{+}(p)}{2\ p}\ \frac{n_F^{-}(q)-n_F^{+}(q)}{2\ q}
\ \frac{p\ q}{r^4} = \frac{T^2}{(4\pi)^2}\frac{1}{3}\left[(1-3\zeta(3))\ \hat\mu^2 -20(\zeta(3)-
\zeta(5))\ \hat\mu^4 + {\cal O}\left(\hat\mu^6\right)\right].
\label{th-2s4}
\ee
Thermal integrals containing the triangle function:

\be
\int_{\bf pq}\frac{n_F^{-}(p)-n_F^{+}(p)}{2p}\frac{n_F^{-}(q)
-n_F^{+}(q)}{2q}\frac{2p\ q}{\Delta(p,q,r)}= \frac{T^2}{(4\pi)^2}\left(\frac{\Lambda}
{4\pi T}\right)^{4\epsilon}\left[\frac{\hat\mu^2}{\epsilon}+ 2(4\ln2+2\gamma+1)
\ \hat\mu^2-\frac{28}{3}\zeta(3)\ \hat\mu^4+{\cal O}\left(\hat\mu^6\right)\right].
\nn
\label{th-2s1}
\ee


Thermal integrals containing both the triangle function and HTL average are listed below:
\be
\int\limits_{\bf pq}\frac{n_F(p)n_F(q)}{4pq}{\mbox Re}\left\langle c^2
\frac{r^2c^2-p^2-q^2}{\Delta(p+i\varepsilon,q,rc)}\right\rangle_c 
   = \frac{T^2}{(4\pi)^2}\left[0.014576+0.238069\ \hat\mu^2 + 0.825164
      \ \hat\mu^4+{\cal O}\left(\hat\mu^6\right)\right].
\label{f1}
\ee

\be
\int\limits_{\bf pq}\frac{n_F(p)n_F(q)}{4pq}{\mbox Re}\left\langle c^4
\frac{r^2c^2-p^2-q^2}{\Delta(p+i\varepsilon,q,rc)}\right\rangle_c
      =  \frac{T^2}{(4\pi)^2}\left[0.017715 + 0.28015\ \hat\mu^2 + 
       0.87321\ \hat\mu^4+{\cal O}\left(\hat\mu^6\right)\right].
\label{f2}
\ee

\be
\int\limits_{\bf pq}\frac{n_F(p)n_F(q)}{4pq}{\mbox Re}\left\langle \frac{q^2}
{r^2}c^2\frac{r^2c^2-p^2-q^2}{\Delta(p+i\varepsilon,q,rc)}\right\rangle_c
      = -\frac{T^2}{(4\pi)^2}\left[0.01158 + 0.17449\ \hat\mu^2 + 0.45566\ 
      \hat\mu^4 + {\cal O}\left(\hat\mu^6\right)\right].
\label{f3}
\ee

\be
\int\limits_{\bf pq}\frac{n_B(p)n_F(q)}{2p q}{\mbox Re}\left\langle \frac{p^2-q^2}
 {r^2}\frac{r^2c^2-p^2-q^2}{\Delta(p+i\varepsilon,q,rc)}\right\rangle_c
    = \frac{T^2}{(4\pi)^2}\left[0.17811 + 1.43775\ \hat\mu^2 - 2.45413\ 
     \hat\mu^4 + {\cal O}\left(\hat\mu^6\right)\right].
\label{f4}
\ee


Second set of integrals involve the variables $r_c=|{\bf p+q}/c|$:
 \be
\int\limits_{\bf pq}\frac{n_F(p)n_B(q)}{2p q}{\mbox Re}\left\langle c^{-1+2\epsilon}
\frac{r_c^2-p^2-q^2}{\Delta(p+i\varepsilon,q,r_c)}\right\rangle_c
     = \frac{T^2}{(4\pi)^2}\left[0.19678 + 1.07745\ \hat\mu^2 - 2.63486\ \hat\mu^4
     + {\cal O}\left(\ \hat\mu^6\right)\right].
\label{f5}
\ee

\be
\int\limits_{\bf pq}\frac{n_F(p)n_B(q)}{2p q}{\mbox Re}\left\langle c^{1+2\epsilon}
\frac{r_c^2-p^2-q^2}{\Delta(p+i\varepsilon,q,r_c)}\right\rangle_c
   = \frac{T^2}{(4\pi)^2}\left[0.048368 + 0.23298\ \hat\mu^2 - 0.65074\ \hat\mu^4
   + {\cal O}\left(\ \hat\mu^6\right)\right].
\label{f6}
\ee

\be
\int\limits_{\bf pq}\frac{n_F(p)n_B(q)}{2p q}\frac{p^2}{q^2}{\mbox Re}\left\langle 
c^{1+2\epsilon}\frac{r_c^2-p^2-q^2}{\Delta(p+i\varepsilon,q,r_c)}\right\rangle_c
   = \frac{T^2}{(4\pi)^2}\left(\frac{\Lambda}{4\pi T}\right)^{4\epsilon}\frac{1}
   {96}\left[\left(1+12\ \hat\mu^2\right)\frac{1}{\epsilon}
\right.\nn\left.
   +\left(7.77236+81.1057\ \hat\mu^2-48.5858\ \hat\mu^4+{\cal O}
   \left(\hat\mu^6\right)\right)\right].
\label{f7}
\ee

\be
\int\limits_{\bf pq}\frac{n_F(p)n_B(q)}{2p q}{\mbox Re}\left\langle c^{1+2\epsilon}
\frac{r_c^2}{q^2}\frac{r_c^2-p^2-q^2}{\Delta(p+i\varepsilon,q,r_c)}\right\rangle_c
    = \frac{T^2}{(4\pi)^2}\left(\frac
    {\Lambda}{4\pi T}\right)^{4\epsilon}\frac{11-8\ln2}{288}\left[\frac{1}{\epsilon}
    \left(1+12\ \hat\mu^2\right)
\right.\nn\left.
   +\left(7.7995 + 70.5162\ \hat\mu^2 - 57.9278\ \hat\mu^4 + {\cal O}
    \left(\ \hat\mu^6\right)\right)\right].
\label{f8}
\ee

\be
\int\limits_{\bf pq}\frac{n_F(p)n_F(q)}{4p q}{\mbox Re}\left\langle c^{-1+2\epsilon}
\frac{r_c^2-p^2}{q^2}\frac{r_c^2-p^2-q^2}{\Delta(p+i\varepsilon,q,r_c)}\right\rangle_c
    = -\frac{T^2}{(4\pi)^2}\left(\frac{\Lambda}{4\pi T}\right)^{4\epsilon}\frac{1}{24}
    \left[\left(1+12\ \hat\mu^2\right)\frac{1}{\epsilon^2}
\right.\nn\left.
    +\frac{2}{\epsilon}\left(1+\gamma+\ln2+ \frac{\zeta'(-1)}{\zeta(-1)} + (24\gamma
    + 48\ln2-7\zeta(3))\ \hat\mu^2 + (31\zeta(5)-98\zeta(3))\ \hat\mu^4 +
     {\cal O}\left(\hat\mu^6\right)\right)
\right.\nn\left.
    + \left(40.3158 + 261.822\ \hat\mu^2 - 1310.69\ \hat\mu^4 + {\cal O}\left(\hat
    \mu^6\right)\right)\right].
\label{f9}
\ee

\be
\int\limits_{\bf pq}\frac{n_B(p)n_F(q)}{2p q}&&{\mbox Re}\left\langle c^{-1+2\epsilon}
\frac{r_c^2-p^2}{q^2}\frac{r_c^2-p^2-q^2}{\Delta(p+i\varepsilon,q,r_c)}\right\rangle_c 
     = -\frac{T^2}{(4\pi)^2}\left(\frac{\Lambda}{4\pi T}\right)^{4\epsilon}\frac{1}{12}
    \left[\frac{1}{\epsilon^2}
\right.\nn&+&\left.
     \frac{1}{\epsilon}\left(2+2\gamma+4\ln2+2\frac{\zeta'(-1)}{\zeta(-1)}-14\zeta(3)\ 
     \hat\mu^2 + 62\zeta(5)\ \hat\mu^4 + {\cal O}\left(\hat\mu^6\right)\right)
\right.\nn &+&\left.
     \left(52.953 - 190.103\ \hat\mu^2 + 780.921\ \hat\mu^4 +
     {\cal O}\left(\hat\mu^6\right)\right)\right]. \hspace{5cm}
\label{f10}
\ee
The integral (\ref{f2}) can be evaluated directly in three dimensions at finite chemical 
potential. The other  integrals Eqs.~(\ref{f3})--(\ref{f10}) can be evaluated following the same
procedure as discussed in \cite{andersen2} at finite chemical potential.


\end{document}